\documentclass[11pt]{article}

\usepackage{amsfonts,amssymb,amsmath}
\usepackage{subfigure}
\usepackage{float}
\usepackage{graphicx}

\begin{document}
\baselineskip=15.5pt \pagestyle{plain} \setcounter{page}{1}

\begin{titlepage}

\vskip 0.8cm

\begin{center}

{\LARGE Holographic thermalization with a chemical potential from Born-Infeld electrodynamics}	
\vskip .3cm

\vskip 1.cm

{\large {Giancarlo Camilo{$^\dagger$}, Bertha Cuadros-Melgar{$^\ddagger$} and Elcio Abdalla{$^\dagger$}}}

\vskip 1.cm

{{$^\dagger$}{\it Instituto de F\'isica, Universidade de S\~ao Paulo\\C.P. 66318, CEP: 05315-970, S\~ao Paulo, Brasil}\\
{$^\ddagger$}{\it Escola de Engenharia de Lorena, Universidade de S\~ao Paulo,\\Estrada Municipal do Campinho S/N, CEP: 12602-810, Lorena, Brasil}} 

\let\thefootnote\relax\footnotetext{Email: {\tt gcamilo@usp.br, bertha@usp.br, eabdalla@usp.br}}
\vspace{1.cm}

{\bf Abstract}

\end{center}

The problem of holographic thermalization in the framework of Einstein gravity coupled to Born-Infeld nonlinear electrodynamics is investigated. We use equal time two-point correlation functions and expectation values of Wilson loop operators in the boundary quantum field theory as probes of thermalization, which have dual gravity descriptions in terms of geodesic lengths and minimal area surfaces in the bulk spacetime. The full range of values of the chemical potential per temperature ratio $\mu/T$ on the boundary is explored.
The numerical results show that the effect of the charge on the thermalization time is similar to the one obtained with Maxwell electrodynamics, namely the larger the charge the later thermalization occurs. The Born-Infeld parameter, on the other hand, has the opposite effect: the more nonlinear the theory is, the sooner it thermalizes. We also study the thermalization velocity and how the parameters affect the phase transition point separating the thermalization process into an accelerating phase and a decelerating phase.

\noindent

\end{titlepage}

\newpage

\tableofcontents

\vfill

\newpage

%%%%%%%%%%%%%%%%%%%%%%%%%%%%%%%%%%%%%%%%%%%%%%%%%%%%%%%%%%%%%%%%%%%%%%%
\section{Introduction}\label{intro}
%%%%%%%%%%%%%%%%%%%%%%%%%%%%%%%%%%%%%%%%%%%%%%%%%%%%%%%%%%%%%%%%%%%%%%%

Over the last years the AdS/CFT correspondence \cite{adscft1,adscft2,adscft3} has proved to be a powerful tool to describe gauge theories at strong coupling regime, where standard perturbative methods fail to work, by mapping them to a dual gravitational theory in higher dimensions whose treatment is tipically much easier. For that reason, it has found a wide range of applications in different areas of theoretical physics going from condensed matter systems to quantum chromodynamics. In particular, if the strongly coupled gauge theory is at zero temperature, it is well known that the dual gravity description involves a purely AdS space, while a finite temperature field theory requires the presence of an asymptotically AdS black hole. An interesting implication of this equivalence will be investigated in the present work, namely, the process of black hole formation in AdS space as the analog of nonequilibrium dynamics that leads a system towards thermal equilibrium after a sudden injection of energy.

One of the most interesting scenarios where these ideas have been applied is to describe properties of the quark-gluon plasma (QGP) formed in heavy ion colliders such as the RHIC and LHC. Recent results suggest that the QGP behaves as an ideal fluid with a very small shear viscosity over entropy density ratio ($\eta/s$) \cite{QGP}. This implies that the QGP takes place at a strong coupling regime and therefore is amenable to a dual gravity treatment. Indeed, holographic calculations using the prototypical $\mathcal{N}=4$ SU($N$) supersymmetric Yang-Mills (SYM) theory at finite $T$ and its string theoretical AdS gravity dual show that there seems to be a small and universal lower limit for the ratio $\eta/s$ for all theories with gravity duals \cite{son1,son2}. This is one of the most prominent predictions of AdS/CFT at the moment. However, while the near-equilibrium dynamics (e.g., transport coefficients such as viscosity and electrical conductivity \cite{conductivity} and more general aspects of dissipative hydrodynamics \cite{fluidgravity,qnm}) of the QGP is well known, the far from equilibrium process of formation of QGP after a heavy ion collision, often referred to as thermalization, is not well understood. The thermalization time scale observed at RHIC is considerably shorter than expected according to perturbative techniques \cite{RHIC}, reinforcing the need of a strong coupling description of the thermalization process.

Some attempts were made in the literature to address this problem from a holographic point of view as a dual process of black hole formation via gravitational collapse in AdS space \cite{danielsson,janik1,janik2,chesler,bhattacharya,pedraza,garfinkle,garfinkle2,lin}. A slightly different and simpler model for holographic thermalization was introduced by Balasubramanian et al. in \cite{balasubramanian,balasubramanian2}, which despite its simplicity captures many important features of the thermalization process. It consists in the collapse of a thin shell of matter described by an AdS-Vaidya metric that interpolates between pure AdS space at early times and Schwarzschild-AdS black hole at late times. The authors used the dynamical background of the collapsing shell to study the time evolution of nonlocal thermalization probes of the boundary conformal field theory with well known dual gravity descriptions in terms of geometric quantities. They considered equal-time two-point correlation functions of local gauge invariant operators, expectation values of Wilson loop operators, and entanglement entropy which correspond in the gravity side to minimal lengths, areas, and volumes in AdS space, respectively. They found that the thermalization is a top-down process (i.e., UV modes thermalize first while IR modes thermalize later), in contrast to the predictions of bottom-up thermalization from perturbative approaches \cite{bottomup}. This has a clear and intuitive interpretation from the AdS/CFT perspective: UV modes correspond to small distance scales in the boundary of AdS space and, therefore, they do not capture much of the details of the collapse process happening deep into the bulk. IR modes, on the other hand, penetrate deeper into the bulk and for that reason are naturally more sensible to details of the bulk dynamics, therefore they should thermalize later. In addition, the authors found that the thermalization time scales typically as $t_{\textrm{therm}}\sim\ell/2$, where $\ell$ is the characteristic length of the probe.

A natural extension of this model was proposed in \cite{galante} (see also \cite{kundu}) to include the effect of a non-vanishing chemical potential $\mu$, which is usually the case in real heavy ion collision processes. The authors considered the collapse of a thin shell of charged matter in the bulk described by an AdS-Vaidya-like metric leading to a thermal equilibrium configuration given by a Reissner-Nordstr\"om-AdS black hole. They argue that varying the charge of the final state black hole from zero to the extremal value allows to explore the full range of chemical potential per temperature ratio $\mu/T$ in the dual strongly coupled QGP. Their main conclusion was that as the charge is increased, the thermalization time for renormalized geodesic lengths and minimal area surfaces becomes larger. Further investigations were made later to include Gauss-Bonnet higher curvature corrections \cite{chinesesGB,chinesesGBQ}, angular momentum \cite{arefeva1}, noncommutative \cite{chinesesNC}, Lifshitz and hyperscaling violating geometries \cite{lifshitzReza,lifshitz}, de Sitter boundary field theories \cite{bin}, as well as more elaborated dynamics for the collapsing shell \cite{baron1,baron2,balasubramanian3} and a discussion on spectral functions of boundary two-point correlators \cite{balasubramanian4}. Related work can also be found in \cite{abajo,abajo2,albash,keranen,hubeny,stricker,ebrahim}.

In this paper we propose a further study of holographic thermalization with a chemical potential using Einstein gravity coupled to Born-Infeld (BI) nonlinear electrodynamics in the bulk. This is a natural generalization of the discussion initiated in \cite{galante}, since it accomodates more elaborated dynamics for the gauge field including (all order) higher-derivatives of $A_\mu$ and, therefore, it may give rise to interesting effects on the chemical potential of the dual boundary theory that are not captured by the Maxwell description. Although BI electrodynamics has its origin a long time ago \cite{BIoriginal} as an attempt to obtain a finite self-energy of point-like charged particles, currently a renewed interest has been raised due to recent developments in superstring theory. In particular, it is well known that the low energy behavior of the vector modes of open strings is governed by the BI action \cite{BIstring1,BIstring2}, while the low energy dynamics of D-branes is given by a similar non-Abelian version of the BI action \cite{BIstring3} (see also \cite{BIstring4,BIstring5}). Consequently, BI electrodynamics provides a promising scenario to explore deviations from Maxwell electrodynamics, specially from the point of view of AdS/CFT calculations where string theory plays a prominent role (see \cite{BIthermo1,BIthermo2,BIhydro1,BIhydro2,BIholosup1,BIholosup2,BIholosup3,BIholosup4,BIholosup5,BIqnm} for an incomplete list of previous works in this direction).

The paper is structured as follows. In section \ref{EBI} we review the black hole solutions of Einstein-Born-Infeld theory in AdS space and construct their Vaidya-like extensions modelling the collapse of a thin charged shell, to be used in the sequence. Section \ref{HT} is devoted to the holographic setup for the non-local observables chosen as probes of thermalization, namely, the equal-time two-point correlators and the expectation value of Wilson loops. In Section \ref{numerics} we give details of the numerical calculations and present all the results with the effects of the chemical potential and BI parameter on the thermalization curves and velocities. Finally, Section \ref{conclusions} contains our concluding remarks.

%%%%%%%%%%%%%
\section{Vaidya AdS Black Hole solutions in Einstein-Born-Infeld theory}\label{EBI}
%%%%%%%%%%%%%

The starting point is the $(d+1)$-dimensional Einstein gravity action with a negative cosmological constant $\Lambda=-d(d-1)/2l^2$ (being $l$ the AdS curvature radius) minimally coupled to Born-Infeld electrodynamics 
\begin{equation}\label{eq:EBIaction}
 S = \frac{1}{16\pi G}\int d^{d+1}x \sqrt{-g}\big[R-2\Lambda+L_{BI}(F)\big]\ ,
\end{equation}
where $L_{BI}(F)$ is given by 
\begin{equation}
 L_{BI}(F) = 4\beta^2\left(1-\sqrt{1+\frac{F_{\mu\nu}F^{\mu\nu}}{2\beta^2}}\right)\ .
\end{equation}
The constant $\beta$ is the BI parameter with dimension of mass.\footnote{In the context of string theory, $\beta$ appears tipically in terms of the string parameter $\alpha'$ via $\beta=1/2\pi\alpha'$.} It is defined in such a way that the limit $\beta\rightarrow\infty$ corresponds to the standard Maxwell Lagrangian. We choose units in which $16\pi G=1$, $G$ being the Newton's constant in $(d+1)$ dimensions.

The charged black hole solution to the equations of motion coming from the action (\ref{eq:EBIaction}), first obtained in \cite{EBI1} (see also \cite{EBI2}), reads
\begin{equation}\label{eq:metricBH}
 ds^2 = -V(r)dt^2+\frac{dr^2}{V(r)}+r^2 d\Omega_{d-1}^2\ ,
\end{equation}
where $d\Omega_{d-1}^2$ denotes the metric on the unit sphere $\mathbb{S}^{d-1}$ and
\begin{eqnarray}\label{eq:V(r)}
 V(r)&=& 1-\frac{M}{r^{d-2}}+\left[\frac{4\beta^2}{d(d-1)}+\frac{1}{l^2}\right]r^2-\frac{2\sqrt{2}\beta}{d(d-1)r^{d-3}}\sqrt{2\beta^2r^{2d-2}+(d-1)(d-2)Q^2}\nonumber\\
 &&+\frac{2(d-1)Q^2}{dr^{2d-4}}\ _{2}F_1\left[\frac{d-2}{2d-2},\frac{1}{2};\frac{3d-4}{2d-2};-\frac{(d-1)(d-2)Q^2}{2\beta^2r^{2d-2}}\right]\ .
\end{eqnarray}
In the above equation $_2F_1(a,b;c;x)$ is the hypergeometric function and $M,Q$ are integration constants related to the ADM mass $\tilde{M}$ and charge $\tilde{Q}$ of the black hole via \footnote{For simplicity, we will keep referring to $M$ and $Q$ hereinafter simply as \lq\lq mass\rq\rq and \lq\lq charge\rq\rq parameters of the black hole without any risk of confusion.}
\begin{eqnarray*}
 \tilde{M}&=&(d-1)\omega_{d-1}M\ ,\\
 \tilde{Q}&=&2\sqrt{2(d-1)(d-2)}\omega_{d-1}Q\ ,
\end{eqnarray*}
$\omega_{d-1}$ being the volume of the $\mathbb{S}^{d-1}$. There is also a purely electric gauge field given by 
\begin{equation}\label{eq:At}
 A =\left(-\sqrt{\frac{d-1}{2(d-2)}}\frac{Q}{r^{d-2}}\ _{2}F_1\left[\frac{d-2}{2d-2},\frac{1}{2};\frac{3d-4}{2d-2};-\frac{(d-1)(d-2)Q^2}{2\beta^2r^{2d-2}}\right]+\Phi\right)dt\ ,
\end{equation}
where $\Phi$ is a constant corresponding to the electrostatic potential at $r\rightarrow\infty$, which will be related to the chemical potential in the dual gauge theory according to the AdS/CFT correspondence. It is defined such that the gauge field vanishes at the horizon, i.e., 
\begin{equation}\label{eq:Phi}
  \Phi=\sqrt{\frac{d-1}{2(d-2)}}\frac{Q}{r_h^{d-2}}\ _{2}F_1\left[\frac{d-2}{2d-2},\frac{1}{2};\frac{3d-4}{2d-2};-\frac{(d-1)(d-2)Q^2}{2\beta^2r_h^{2d-2}}\right]\ .
\end{equation}

The electric field associated to (\ref{eq:At}) is finite at the origin $r=0$, which is a key feature of BI theories. The black hole function (\ref{eq:V(r)}), on the other hand, is in general singular at the origin. Such a singularity is hidden behind an event horizon provided the free parameters are chosen so that the equation $V(r_h)=0$ admits a real positive solution. We should also mention that taking the limit $\beta\rightarrow\infty$ in (\ref{eq:V(r)}) gives the well known Reissner-Nordstr\"om-AdS black hole studied in \cite{emparanRN}.

The solution (\ref{eq:metricBH}) has the topology of $\mathbb{R}\times\mathbb{S}^{d-1}$ at the AdS boundary $r\rightarrow\infty$. In the context of the AdS/CFT correspondence it is interesting to consider the limit where the AdS boundary is $\mathbb{R}^{1,d-1}$ instead, since one is often interested in dual gauge theories living on flat space. This procedure is known in the literature as the \lq\lq infinite volume limit\rq\rq, and it arises only due to the presence of a negative cosmological constant \cite{emparanRN}. The idea is to introduce a dimensionless parameter $\lambda$ (which will be set to $\infty$) and rescale all dimensionful quantities as $r\rightarrow \lambda^{1/d}r,t\rightarrow \lambda^{-1/d}t,M\rightarrow \lambda M,Q\rightarrow \lambda^{(d-1)/d}Q,\beta\rightarrow \beta,l\rightarrow l$ while at the same time blow up the $\mathbb{S}^{d-1}$ as $l^2d\Omega_{d-1}^2\rightarrow\lambda^{-2/d}\sum_{i=1}^{d-1}dx_i^2$. This leaves the $(t,r)$ block of the metric almost invariant (except for the contribution of the constant term in (\ref{eq:V(r)})). Finally, taking $\lambda\rightarrow\infty$ yields 
\begin{equation}\label{eq:metricBB}
  ds^2 = -U(r)dt^2+\frac{dr^2}{U(r)}+ \frac{r^2}{l^2}\sum_{i=1}^{d-1}dx_i^2\ ,
\end{equation}
where $U(r) \equiv V(r)-1$. Notice that now the horizon, defined by $U(r_h)=0$, is planar instead of spherical, so we should refer to (\ref{eq:metricBB}) as a black brane instead of a black hole. In order to avoid the coordinate singularity at $r=r_h$ it will be interesting to express the metric in Eddington-Finkelstein coordinates by introducing a new time coordinate $v$ defined by $dv=dt+dr/U(r)$, and also it will be convenient to work with an inverse radial coordinate $z=l^2/r$ such that the AdS boundary stays at $z=0$ while the singularity $r=0$ sits at infinity. The resulting metric is
\begin{equation}\label{eq:metricBBEF}
  ds^2 = \frac{l^2}{z^2}\left[-f(z)dv^2-2dvdz+\sum_{i=1}^{d-1}dx_i^2\right]\ ,
\end{equation}
where we have defined
\begin{equation}
 f(z) = \frac{z^2}{l^2}\left[V\left(\frac{l^2}{z}\right)-1\right]\ .
\end{equation}
Notice that $f(z)\rightarrow 1$ near the AdS boundary $z=0$.

The Hawking temperature of a black hole in the context of AdS/CFT can be viewed as the equilibrium temperature of the dual field theory living on the boundary. It is obtained as usual by continuing the black hole metric to its Euclidean version via $t =-it_E$ and demanding the absence of conical singularities at the horizon. This results in a periodic Euclidean time $t_E$ whose period is identified with the inverse Hawking temperature. For the AdS Einstein-Born-Infeld black brane (\ref{eq:metricBB}) this calculation gives
\begin{equation}\label{eq:THawking}
 T = \frac{1}{4\pi r_h}\left[\left(\frac{4\beta^2}{d-1}+\frac{d}{l^2}\right)r_h^2-\frac{2\sqrt{2}\beta}{(d-1)r_h^{d-3}}\sqrt{2\beta^2r_h^{2d-2}+(d-1)(d-2)Q^2}\right]\ .
\end{equation}
This expression reduces to the Hawking temperature of the Reissner-Nordstr\"om-AdS black hole in the Maxwell limit $\beta\rightarrow\infty$ \cite{galante}. When $T=0$, the black brane is called extremal. If we think of all the parameters but the charge as fixed, then we can characterize the extremal black brane solution by a maximal value of charge $Q_{ext}$ given by
\begin{equation}\label{eq:Qext}
 Q_{ext}^2 = \frac{d}{(d-2)l^2}\left[1+\frac{d(d-1)}{8l^2\beta^2}\right]r_h^{2d-2}\ .
\end{equation}

According to the AdS/CFT dictionary, the asymptotic value of the time component $A_t(r)$ of the gauge field at the AdS boundary $r\rightarrow\infty$ (namely, the constant $\Phi$ in equation (\ref{eq:At})) corresponds to the chemical potential $\mu$ in the dual quantum field theory, $\mu\sim\lim_{r\rightarrow\infty}A_t(r)$. Actually, the precise relation should include some scale $\xi$ with length units since the chemical potential must have energy units (or [length]$^{-1}$) while $A_\mu$ as defined by the action (\ref{eq:EBIaction}) is dimensionless. Hence, the chemical potential per temperature ratio of the boundary field theory is given by 
\begin{equation}\label{eq:chempot}
\frac{\mu}{T} =\frac{1}{T} \lim_{r\rightarrow\infty}\frac{A_t(r)}{\xi} = \frac{\Phi}{\xi T}\ , 
\end{equation}
with $\Phi$ and $T$ given by expressions (\ref{eq:Phi}) and (\ref{eq:THawking}), respectively. A remarkable feature is that if the horizon radius $r_h$ and the BI parameter $\beta$ are kept fixed, then by varying the charge from $Q=0$ (vanishing $\Phi$) to $Q=Q_{ext}$ (vanishing $T$) it is possible to explore the whole range of values of the ratio $\mu/T$ in the dual field theory, i.e., from $\mu/T=0$ to $\infty$.

A Vaidya-like extension of the BI AdS black brane metric (\ref{eq:metricBB}) can be constructed by promoting the mass and charge to arbitrary functions $M(v)$ and $Q(v)$ of the advanced time $v$. The resulting dynamical metric has the same form as in (\ref{eq:metricBBEF}) but now with $f(v,z)$ instead of $f(z)$ due to the time dependence introduced on the mass and charge. The same also holds for the gauge field (\ref{eq:At}), which now becomes $A_\mu(v,z)$. Such a spacetime describes the collapse of a thin-shell of charged dust from the boundary of the AdS space towards the bulk interior.

Of course such a metric is not a solution of the action (\ref{eq:EBIaction}) anymore: there must be some external matter action $S_{m}$ sourcing the time variation of $M(v)$ and $Q(v)$. If we take this external contribution into account, the Einstein-BI equations of motion become (we restore the factors of $G$ for a moment):
\begin{align}
 \!\!\!R_{\mu\nu}-\frac{1}{2}Rg_{\mu\nu}+\Lambda g_{\mu\nu}-2\beta^2g_{\mu\nu}\left(1-\sqrt{1+F^2/2\beta^2}\right)-\frac{2F_{\mu\alpha}F_{\nu}^{\ \alpha}}{\sqrt{1+F^2/2\beta^2}} &= -8\pi GT_{\mu\nu}^{(m)}\\
 \nabla_\mu\left(\frac{F^{\mu\nu}}{\sqrt{1+F^2/2\beta^2}}\right) &= -8\pi GJ^{\nu}_{(m)}\!\ .
\end{align}
The Vaidya-BI-AdS metric above-mentioned is a solution to these equations provided the external sources satisfy
\begin{align}
 8\pi G T^{(m)}_{\mu\nu} &=\frac{(d-1)z^{d-1}}{2l^{2d-2}}\left[\dot{M}(v)-2\left(\frac{z}{l^2}\right)^{d-2}\!\!\!_{2}F_1\big[{\scriptstyle\frac{d-2}{2d-2},\frac{1}{2};\frac{3d-4}{2d-2};-\frac{(d-1)(d-2)Q(v)^2}{2\beta^2(z/l^2)^{2-2d}}}\big]Q(v)\dot{Q}(v)\right]\delta_\mu^v\delta_\nu^v\\
 8\pi G J_{(m)}^\nu &= \sqrt{\frac{(d-1)(d-2)}{2}}\frac{z^{d+1}}{l^{d+2}}\dot{Q}(v)\delta^\nu_z\ ,
\end{align}
where the dot denotes $\partial_v$. We notice that there is no $\beta$ dependence on $J_{(m)}^\nu$ above, and indeed this is exactly the same current found in \cite{galante} in the Vaidya-Reissner-Nordstr\"om-AdS case. $T^{(m)}_{\mu\nu}$, on the other hand, differs from the corresponding one in the Reissner-Nordstr\"om case due to the hypergeometric term (but naturally reduces to it in the $\beta\rightarrow\infty$ limit, since $_2F_1[a,b;c;0]=1$).

%%%%%%%%%%%%%%%%%%%%%%%%%%%%%%%%%%%%%
\section{Holographic thermalization}\label{HT}
%%%%%%%%%%%%%%%%%%%%%%%%%%%%%%%%%%%%%

In this section we are going to study the thermalization process of a strongly coupled quantum field theory whose bulk gravity dual corresponds to the Einstein-Born-Infeld system presented in the previous section. According to the AdS/CFT correspondence, a zero temperature state on the $d$ dimensional boundary theory is dual to pure $AdS_{d+1}$ in the bulk, while a thermal state corresponds to the BI-AdS black brane (\ref{eq:metricBBEF}). Therefore, any dynamical bulk spacetime which interpolates between these two situations would be a natural candidate to holographically model the nonequilibrium process leading to thermalization of the boundary theory after a rapid injection of energy. 

Following \cite{balasubramanian}, we choose our dynamical bulk metric to be the Vaidya-BI-AdS metric discussed in Section \ref{EBI}, namely (we set the AdS radius $l=1$ hereafter)
\begin{equation}\label{eq:Vaidya}
  ds^2 = \frac{1}{z^2}\left[-f(v,z)dv^2-2dvdz+\sum_{i=1}^{d-1}dx_i^2\right]\ ,
\end{equation}
with the mass and charge functions given by (see \cite{pedrazaNEC} for an interesting discussion on the corresponding bulk null energy condition)
\begin{align}
M(v)&=\frac{M}{2}\left(1+\tanh\frac{v}{v_0}\right)\label{eq:Mfunc}\\
Q(v)&=\frac{Q}{2}\left(1+\tanh\frac{v}{v_0}\right)\label{eq:Qfunc}\ .
\end{align}
Clearly, for $v\rightarrow-\infty$ we have pure AdS ($M(v)=0=Q(v)$) and for $v\rightarrow\infty$ we have $M(v)=M$ and $Q(v)=Q$ which is the BI-AdS black brane (\ref{eq:metricBBEF}). Indeed, equations (\ref{eq:Mfunc})-(\ref{eq:Qfunc}) are just smooth versions (convenient for the numerical analysis) of the step functions $M(v)=M\theta(v)$ and $Q(v)=Q\theta(v)$, which represent a shock wave (a zero thickness shell of charged matter suddenly forming at $v=0$). The constant $v_0$ represents a finite shell thickness and for $v_0\rightarrow0$ we go back to the step function.

The next step is to choose a set of observables to use as probes of thermalization. Since local observables in the boundary such as expectation values of the energy-momentum tensor are not sensitive to the thermalization process, one needs to consider extended non-local observables.\footnote{Holography provides a geometric intuition for why local operators are insensitive to details of the progress towards thermalization: being local, they are only sensitive to phenomena happening in their vicinity near the AdS boundary. Thus they are not aware of the details of phenomena occurring near the thermal scale. We need observables dual to AdS quantities that probe deeper into the bulk in order to see signals of thermalization.} In this work we shall focus on equal time two-point correlation functions and expectation values of rectangular Wilson loops, which have well known holographic descriptions in the bulk in terms of renormalized geodesic lengths and minimal area surfaces, respectively. A third observable that could be used is the entanglement entropy of boundary regions, which have a very similar description in terms of minimal volumes of codimension-two surfaces in the bulk. However, as the results of entanglement entropy lead essentially to the same conclusions, we will not show them here in order to avoid unnecessary repetitions.

\subsection{Renormalized geodesic lengths and two-point functions}

We start with the equal-time two-point correlation functions of local gauge invariant operators $\mathcal{O}(t,\mathbf{x})$ of conformal dimension $\Delta$. The AdS/CFT correspondence provides a simple geometrical way to compute it in the bulk gravity dual when the operator $\mathcal{O}$ is \lq\lq heavy\rq\rq, i.e., when $\Delta\gg1$ \cite{balasubramanianparticledetection}. Namely,
\begin{equation}\label{eq:geodesicaprox}
 \langle \mathcal{O}(t,\mathbf{x})\mathcal{O}(t,\mathbf{x'})\rangle \approx  e^{-\Delta \mathcal{L}}\ ,
\end{equation}
where $\mathcal{L}$ stands for the length of the bulk geodesic between the points $(t,\mathbf{x})$ and $(t,\mathbf{x'})$ located on the AdS boundary.\footnote{If there is more than one geodesic we should sum over them on the right-hand side.} Actually, one should be careful when doing such an approximation because the geodesic length above is divergent due to the contribution of the AdS boundary (because the AdS metric itself diverges at the boundary $z=0$). In order to extract a meaningful quantity we need to introduce a cutoff $z_0$. It turns out that the divergent part of $\mathcal{L}$ is universal and equals to $2\ln(2/z_0)$ \cite{balasubramanian}. Then, we define a (finite) renormalized geodesic length $\mathcal{L}_{ren}=\mathcal{L}-2\ln(2/z_0)$, which will be related to the renormalized two-point function $\langle \mathcal{O}(t,\mathbf{x})\mathcal{O}(t,\mathbf{x'})\rangle_{ren}$ just as in equation (\ref{eq:geodesicaprox}).

We choose the coordinate axes for the boundary directions $(t,\mathbf{x})$ such that the spatial separation $\ell=|\mathbf{x}-\mathbf{x'}|$ between the points lies entirely over the $x^1$ direction, i.e., we consider space-like geodesics between the boundary points $(t, x^1=-\ell/2,\ldots)$  and $(t, x'^1=+\ell/2,\ldots)$, where the ellipsis denotes the remaining coordinates $(x^2,\ldots,x^{d-1})$ which are the same for both points. Then, by symmetry the geodesics cannot depend on coordinates other than $x^1$, and we can use $x^1$ as the geodesic parameter (we call it simply $x$ hereinafter). The solutions of the geodesic equations are then given by a pair of functions $v(x)$ and $z(x)$. The boundary conditions at the AdS boundary $z=z_0$ are
\begin{equation}\label{eq:bc}
 z(\pm\ell/2)=z_0, \qquad v(\pm\ell/2)=t\ .
\end{equation}

The length functional between the referred points follows immediately from the line element (\ref{eq:Vaidya}) as being
\begin{equation}\label{eq:length} 
L[v,z]= \int_{-\ell/2}^{\ell/2}dx \frac{\sqrt{1-2z'(x)v'(x)-f(v,z)v'(x)^2}}{z(x)}\ .
\end{equation}
It clearly depends on the path taken from one point to the other. The geodesic corresponds to the functions $v(x)$ and $z(x)$ that minimize the length and can be found by standard methods. The geodesic length (which we will call $\mathcal{L}$) is just the value of the length functional $L$ evaluated at the geodesic solution.

The variational problem simplifies by noticing that the integrand in (\ref{eq:length}) does not depend explicitly on $x$ and, therefore, there is a conserved Hamiltonian $\mathcal{H}$ given by
\begin{equation}
 \mathcal{H}=\frac{1}{z(x)\sqrt{1-2z'(x)v'(x)-f(v,z)v'(x)^2}}\ .
\end{equation}
Using the conditions on the turning point of the geodesic,
\begin{equation}\label{eq:bcmod}
 z(0)=z_*,\qquad v(0)=v_*,\qquad z'(0)=0,\qquad v'(0)=0\ ,
\end{equation}
arising from the fact that the geodesic must be symmetric with respect to $x=0$, the conservation equation simplifies to
\begin{equation}\label{eq:conserv}
\sqrt{1-2z'(x)v'(x)-f(v,z)v'(x)^2}=\frac{z_*}{z(x)}\ .
\end{equation}
The advantage of working with the boundary conditions (\ref{eq:bcmod}) instead of the original ones (\ref{eq:bc}) is that we can use the conservation equation above to write the geodesic length (\ref{eq:length}) as
\begin{equation}\label{eq:geolength} 
\mathcal{L}= 2\int_{0}^{\ell/2}dx \frac{z_*}{z(x)^2}\ .
\end{equation}
Hence, after solving the equations of motion and finding the geodesic functions $v(x)$ and $z(x)$ for a given pair of initial conditions $(v_*,z_*)$ it is a trivial task to use the relations (\ref{eq:bc}) to read the corresponding values of boundary separation $\ell$ and time $t$, as well as obtaining the corresponding renormalized geodesic length $\mathcal{L}_{ren}=\mathcal{L}-2\ln(2/z_0)$ by means of expression (\ref{eq:geolength}).

The hard part is to find the geodesic, which means to solve the Euler-Lagrange equations for $z(x)$ and $v(x)$. With the help of the conservation equation (\ref{eq:conserv}) they can be written, respectively, as
\begin{subequations}\label{eq:geoeqs}
\begin{align}
 2-2v'(x)^2f(v,z)-4v'(x)z'(x)-2z(x)v''(x) + z(x)v'(x)^2\partial_zf(v, z) &= 0\\
 z(x)v''(x)f(v,z)+z(x)z''(x)+z(x)z'(x)v'(x)\partial_zf(v,z)+\frac{1}{2}z(x)v'(x)^2\partial_vf(v,z) &= 0\ .
\end{align}
\end{subequations}
This is a set of coupled, highly nonlinear differential equations and for that reason it is quite hard to handle with analitical methods. However, for a given pair $(v_*,z_*)$ it is possible to find a numerical solution subject to the boundary conditions (\ref{eq:bcmod}). Indeed, by solving for sufficiently many pairs of initial conditions $(v_*,z_*)$ (carefully chosen in order to give the same boundary separation $\ell$ and different times), we can track time after time the whole evolution of the geodesics in the Vaidya-BI-AdS spacetime. In particular, if we calculate the renormalized geodesic length of each of these solutions using (\ref{eq:geolength}) we will be able to see the full time evolution of $\mathcal{L}_{ren}$ towards thermalization. This will be done in Section \ref{numerics}, where we provide a detailed explanation of the numerical procedure as well as the choice of parameters and show our results.

\subsection{Minimal area surfaces and Wilson loops}

We now study a second class of thermalization probes, namely the expectation values of Wilson loop operators in the boundary field theory. The Wilson loop is a non-local gauge-invariant observable defined as the path-ordered integral of the gauge field over a closed path $\mathcal{C}$:
\begin{equation}\label{eq:wldef}
 W(\mathcal{C})=\frac{1}{N}\textrm{Tr}\left(\mathcal{P}e^{\oint_\mathcal{C}A_\mu dx^\mu}\right)\ ,
\end{equation}
where $N$ is the number of colors and $A_\mu$ is the non-abelian gauge field. Wilson loops contain useful information about the non-perturbative behavior of non-abelian gauge theories, such as whether they exhibit confinement or not. It is possible, in principle, to express all gauge-invariant functions of $A_\mu$ in terms of Wilson loops by appropriate choices of the path $\mathcal{C}$, but unfortunately they are in general hard to compute.

The AdS/CFT correspondence again provides an elegant way to compute the expectation value of Wilson loops of a strongly coupled gauge theory with a gravitational dual in terms of a geometrical quantity in the bulk \cite{maldacenaWL}:
\begin{equation}\label{eq:wl}
 \langle W(\mathcal{C})\rangle \approx  e^{-\frac{1}{\alpha'}\mathcal{A}(\Sigma_{0})}\ ,
\end{equation}
where $\alpha'$ is the inverse string tension, $\Sigma_0$ denotes the minimal area bulk surface whose boundary is the original contour $\mathcal{C}$, and $\mathcal{A}(\Sigma_{0})$ is the area of that surface. $\Sigma_0$ will be a solution of the bosonic part of the string action (the Nambu-Goto action), which is nothing but the area of the classical world-sheet with $\mathcal{C}$ as its boundary. Indeed, equation (\ref{eq:wl}) has its origin in a saddle-point approximation of the string theory partition function around the classical solution, which is only valid when the dual gauge theory has strong coupling.

Now we proceed to compute the minimal area surfaces in the Vaydia-BI-AdS spacetime as described before just as we did for the geodesic lengths. We will focus on a spacelike rectangular Wilson loop on the boundary. The rectangle $\mathcal{C}$ can always be chosen to lie on the $x^1$-$x^2$ plane, centered at the origin, with sides $\ell$ on the $x^1$ direction and $R$ on the $x^2$ direction. One also assumes translational invariance along $x^2$, such that the shape of the bulk surface depends only on $x^1$ and again we can use $x^1\equiv x$ to parametrize the functions $v(x)$ and $z(x)$ that characterize the surface. The boundary conditions at the AdS boundary $z=z_0$ are again given by equations (\ref{eq:bc}).
%\begin{equation}\label{eq:bc}
% z(\pm\ell/2)=z_0, \qquad v(\pm\ell/2)=t\ .
%\end{equation}

Using the Vaidya-BI-AdS metric (\ref{eq:Vaidya}), the Nambu-Goto action (or area functional divided by $2\pi$) becomes 
\begin{equation}\label{eq:nambu} 
A_{NG}[v,z]=\frac{R}{2\pi}\int_{-\ell/2}^{\ell/2}dx \frac{\sqrt{1-2z'(x)v'(x)-f(v,z)v'(x)^2}}{z(x)^2}\ .
\end{equation}
Notice that an obvious consequence of our assumption of translational invariance is that the length $R$ factorizes. Since we are interested just in the $\ell$ dependence, we can study $A_{NG}/R$ instead of $A_{NG}$ itself and forget about $R$ in what follows. As for the geodesics, the pair of functions $\left(v(x),z(x)\right)$ that minimizes the Nambu-Goto action will be the minimal surface $\Sigma_0$. The on-shell value of the Nambu-Goto action (i.e., $A_{NG}[\Sigma_0]$), which we call $\mathcal{A}$, will be our object of interest. 

The subsequent calculation is closely analogous to the geodesics case. There is again a conserved Hamiltonian associated to (\ref{eq:nambu}) and we can introduce the alternative boundary conditions on the turning point of the minimal surface, which are the same as equations (\ref{eq:bcmod}), to find an expression similar to (\ref{eq:conserv}) for the conservation equation. Replacing this back into (\ref{eq:nambu}), the on-shell Nambu-Goto action becomes
\begin{equation}\label{eq:nambuonshell} 
\mathcal{A}= \frac{R}{\pi}\int_{0}^{\ell/2}dx \frac{z_*^2}{z(x)^4}\ ,
\end{equation}
which can be used to easily obtain the minimal area surface once we have solved the equations of motion and found the functions $v(x)$ and $z(x)$ for a given pair of initial conditions $(v_*,z_*)$. The corresponding values of boundary separation $\ell$ and time $t$ can be read from the original conditions (\ref{eq:bc}) as well. Here again we have to face the problem that the area $\mathcal{A}$ diverges due to the contribution near the AdS boundary, but we can regularize the divergent part using again a cutoff $z_0$ and subtract it from $\mathcal{A}$ to define the renormalized minimal area $\mathcal{A}_{ren}=\mathcal{A}-R/\pi z_0$ \cite{balasubramanian}.

The functions $z(x)$ and $v(x)$ are found by solving the Euler-Lagrange equations coming from the action (\ref{eq:nambu}). The simplified equations of motion (after using the conservation equation) are, respectively
\begin{subequations}\label{eq:areaeqs}
\begin{align}
z(x)v'(x)^2\partial_z f(v,z)-4v'(x)^2f(v,z)-2z(x)v''(x) -8v'(x)z'(x)+4 &= 0\\
v'(x)z'(x)\partial_z f(v,z)+\frac{1}{2}v'(x)^2\partial_v f(v,z)+v''(x)f(v,z)+z''(x) &= 0\ .
\end{align}
\end{subequations}
As before, for a given pair $(v_*,z_*)$ we can solve this set of equations numerically subject to the boundary conditions (\ref{eq:bc}). Thus, for some chosen length $\ell$ of the Wilson loop, by iterating for enough pairs $(v_*,z_*)$ we can track the whole time evolution of the minimal surfaces and, in particular, of their renormalized areas $\mathcal{A}_{ren}$ towards thermalization. This will be the aim of Section \ref{numerics}.

%%%%%%%%%%%%%%%%%%%%%%%%%%%%%%%%%%%%%
\section{Numerical results}\label{numerics}
%%%%%%%%%%%%%%%%%%%%%%%%%%%%%%%%%%%%%

\subsection{Renormalized geodesic lengths}

In this section, we numerically solve the geodesic equations of motion (\ref{eq:geoeqs}) in order to find how the geodesic length evolves with time. Afterwards, we explore how the charge of the black hole and BI parameter affect the thermalization time. For the latter it will be convenient in the numerical calculations to use an inverse BI parameter $b=1/\beta$ instead of the original $\beta$, such that the Maxwell limit is $b\rightarrow 0$ and increasing $b$ accounts for increasingly nonlinear electrodynamics. 

\begin{figure}[t]
	\centering
		\includegraphics[width=0.50\textwidth]{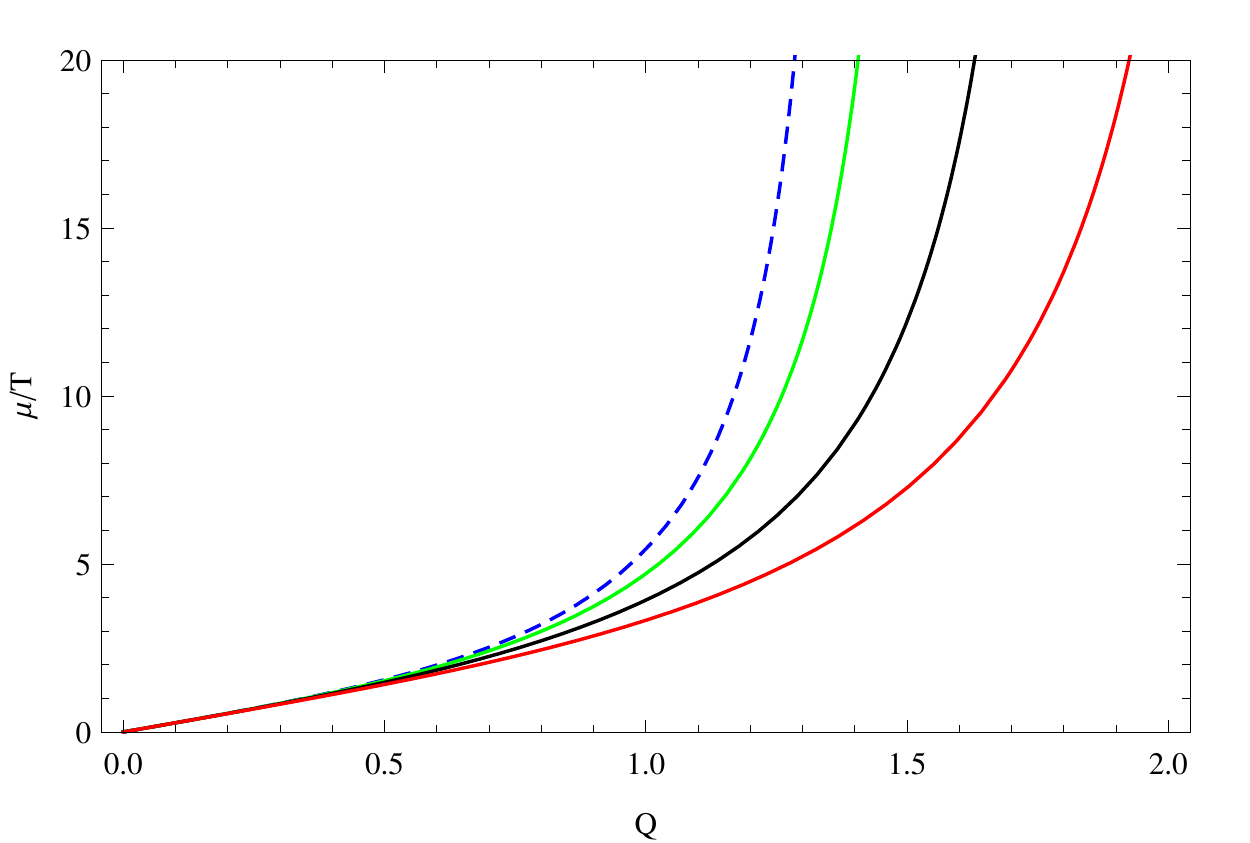}
		\caption{The chemical potential per temperature ratio in the $d=4$ field theory as a function of the charge $Q$ of the black hole for different values of the inverse BI parameter: $b=0$ (dashed blue), $b=0.4$ (green), $b=0.7$ (black), $b=1$ (red). The dashed curve represents the Maxwell limit case.}
	\label{fig:mut}
\end{figure}

First of all, we fix the free parameters. We will take the shell thickness and AdS space UV cut-off to be $v_0 = 0.01$ and $z_0 = 0.01$, respectively. Since the effect of the number of spacetime dimensions and boundary separation on the thermalization probes has already been analyzed in previous works \cite{balasubramanian,balasubramanian2,galante}, we focus here in the case $d=4$ (namely, AdS$_5$ space, which is dual to a $4-$dimensional gauge theory) and a fixed boundary separation $\ell=4$. The mass $M$ of the final state black brane can be expressed in terms of the radius of its event horizon using the definition of $r_h$, i.e., the largest solution of $U(r_h)=0$. Then, if we choose to fix the horizon at $r_h=1$,\footnote{This will be interesting to compare the thermalization for different values of $Q$ and $b$, since the black hole will always form at the same location for all $Q,b$.} the mass is given by
\begin{equation}\label{eq:M}
 M=1+\frac{1}{3b^2}\left(1-\sqrt{1+3b^2Q^2}\right)+\frac{3}{2}\ _2F_1\left[\frac{1}{3},\frac{1}{2};\frac{4}{3};-3b^2Q^2\right]\ ,
\end{equation}
which of course only holds provided that $b$ and $Q$ take values consistent with the existence of an event horizon. For a given $b$ this means that $Q$ is allowed to take values from $Q=0$ to the extremal value (\ref{eq:Qext}), which now reads 
\begin{equation}\label{eq:Qextb}
 Q_{ext}(b) = \sqrt{2+3b^2}\ .
\end{equation}
As we pointed out in Section \ref{EBI}, for each $b$, considering values of charge $0\le Q\le Q_{ext}(b)$ we can study all the range $0\le\mu/T\le\infty$ in the dual gauge theory.
It is instructive to illustrate this fact here by looking at the form of expression (\ref{eq:chempot}) in the present case (we choose the scale $\xi=1$ for simplicity), namely,
\begin{equation}
\frac{\mu}{T} = \frac{3\sqrt{3}\pi b^2Q\ _2F_1\left[\frac{1}{3},\frac{1}{2};\frac{4}{3};-3b^2Q^2\right]}{2+6b^2-2\sqrt{1+3b^2Q^2}}\ .
\end{equation}
A plot of this as a function of the charge for distinct values of $b$ is shown in Figure \ref{fig:mut}, showing that indeed all the range of $\mu/T$ is covered. We also notice that for small values of charge (up to $\sim0.5$) the BI parameter has no effect on the chemical potential since all the curves agree, so we will concern only about charges above this value in what follows.

Now we proceed to investigate the effects of $Q$ and $b$ on the thermalization process. We choose the test values $b=0,0.4,1$ and $Q=0.5,\ldots,Q_{ext}(b)$ for the numerical analysis. The procedure is the following: for a given $(b,Q)$ pair we solve the geodesic equations (\ref{eq:geoeqs}) subject to the boundary conditions (\ref{eq:bcmod}) characterized by a pair of values $(v_*,z_*)$. We do this recursively for various pairs of initial conditions and collect from these just those yielding a boundary separation $\ell=4$.\footnote{There is a small subtlety here. $\ell$ is determined numerically, via equation (\ref{eq:bc}), after we have solved the equations using the modified boundary conditions (\ref{eq:bcmod}), hence we must establish a criterion for what we mean by \lq\lq$\ell=4$\rq\rq. We adopt the convention of $0.0005$ tolerance, meaning that \lq\lq$\ell=4$\rq\rq here corresponds to $\ell\in(3.9995,4.0005)$.} Each of these collected solutions correspond to a different stage of the motion of the geodesics, as we may check by computing the corresponding boundary time $t$ via equation (\ref{eq:bc}). We then calculate the renormalized geodesic length $\mathcal{L}_{ren}$ of each of these collected solutions and construct a list of points $(t,\mathcal{L}_{ren}(t))$ which contains all the information about the time evolution of the renormalized geodesic length. Actually, it will be convenient to divide all the lengths by $\ell$ in order to obtain a dimensionless, $\ell-$independent quantity $\tilde{\mathcal{L}}=\frac{\mathcal{L}_{ren}}{\ell}$. In addition, we subtract from this the final (thermal) value just to force all thermalization curves to end at zero, so that the quantity to be plotted is $\tilde{\mathcal{L}}-\tilde{\mathcal{L}}_{\textrm{thermal}}$ versus $t$. 

\begin{figure}[htbp]% Geodesic profiles
    \centering
    \subfigure[$b=0$ at $t=0.37$]{\includegraphics[height=1.1in]{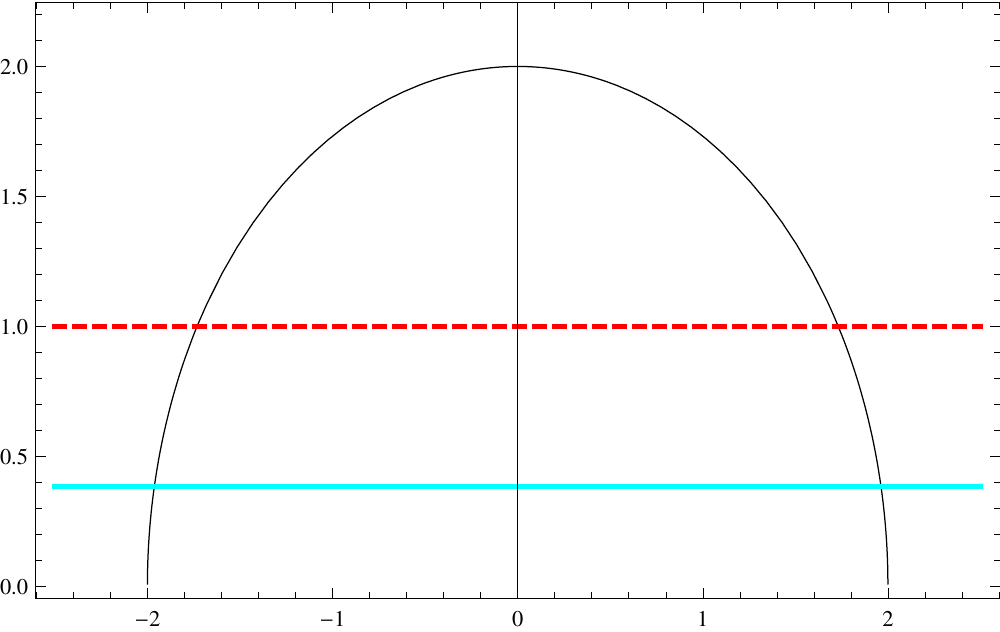}}\qquad
    \subfigure[$b=0.4$ at $t=0.37$]{\includegraphics[height=1.1in]{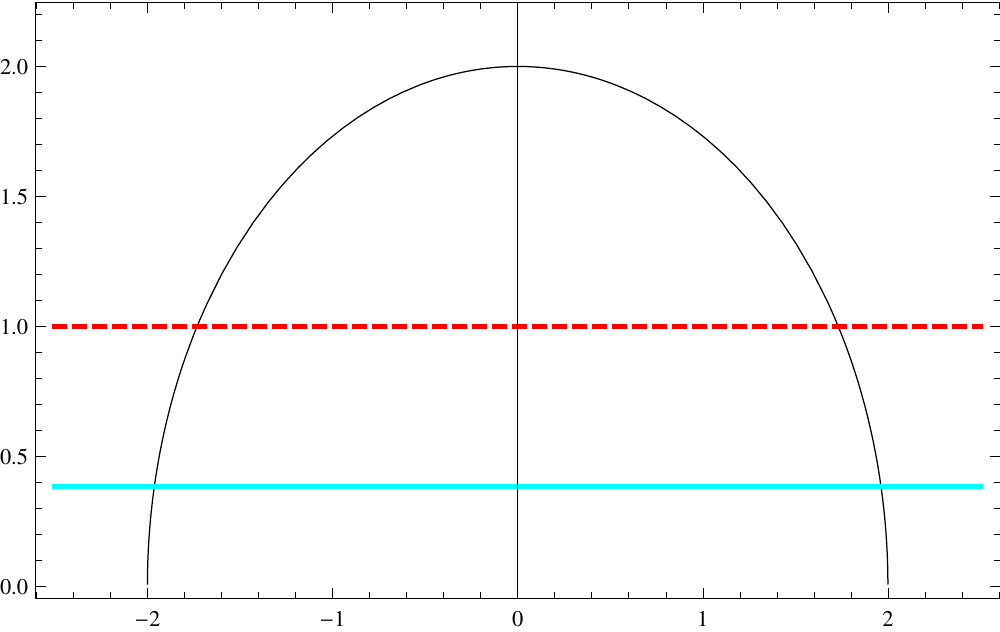}}\qquad
    \subfigure[$b=1$ at $t=0.37$]{\includegraphics[height=1.1in]{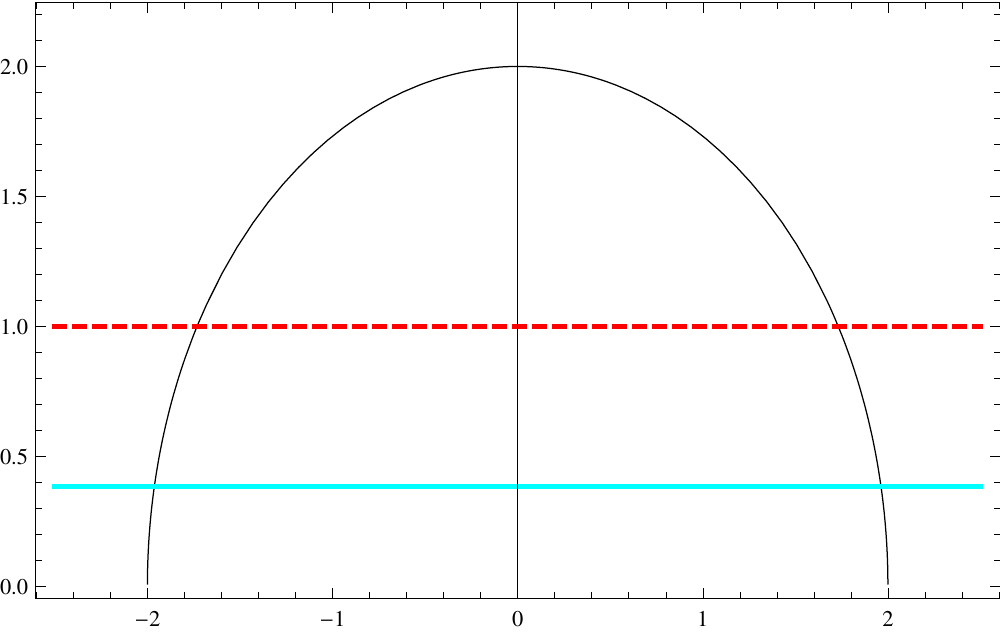}}\\
    \subfigure[$b=0$ at $t=0.82$]{\includegraphics[height=1.1in]{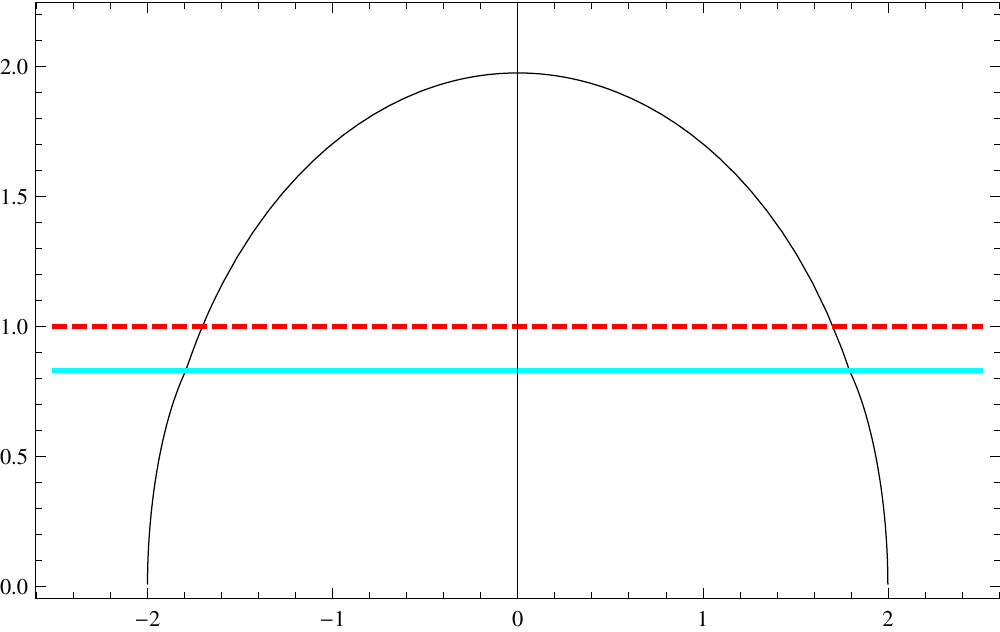}}\qquad
    \subfigure[$b=0.4$ at $t=0.82$]{\includegraphics[height=1.1in]{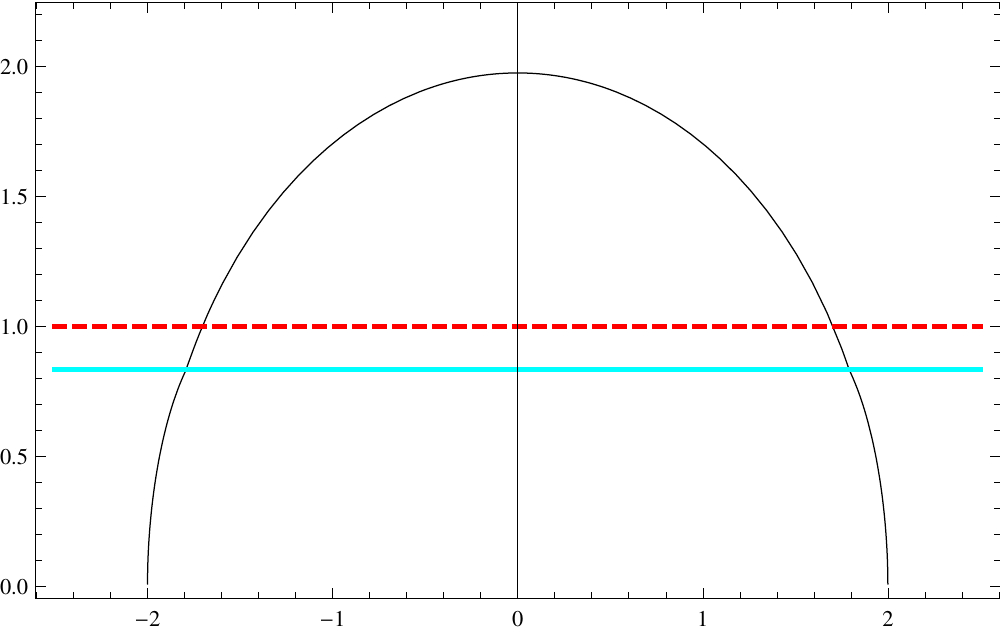}}\qquad
    \subfigure[$b=1$ at $t=0.82$]{\includegraphics[height=1.1in]{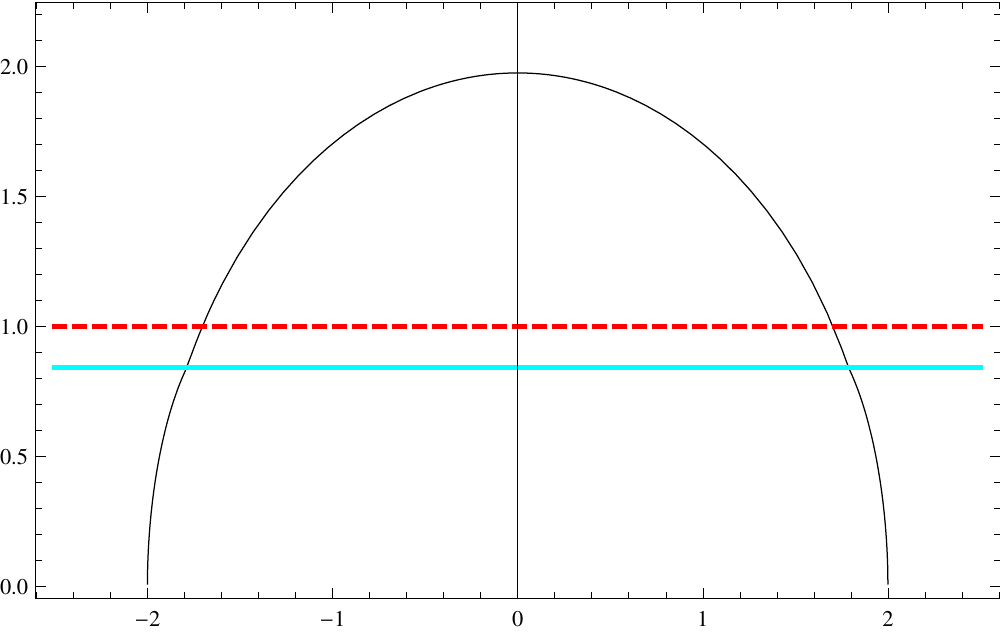}}\\
    \subfigure[$b=0$ at $t=1.54$]{\includegraphics[height=1.1in]{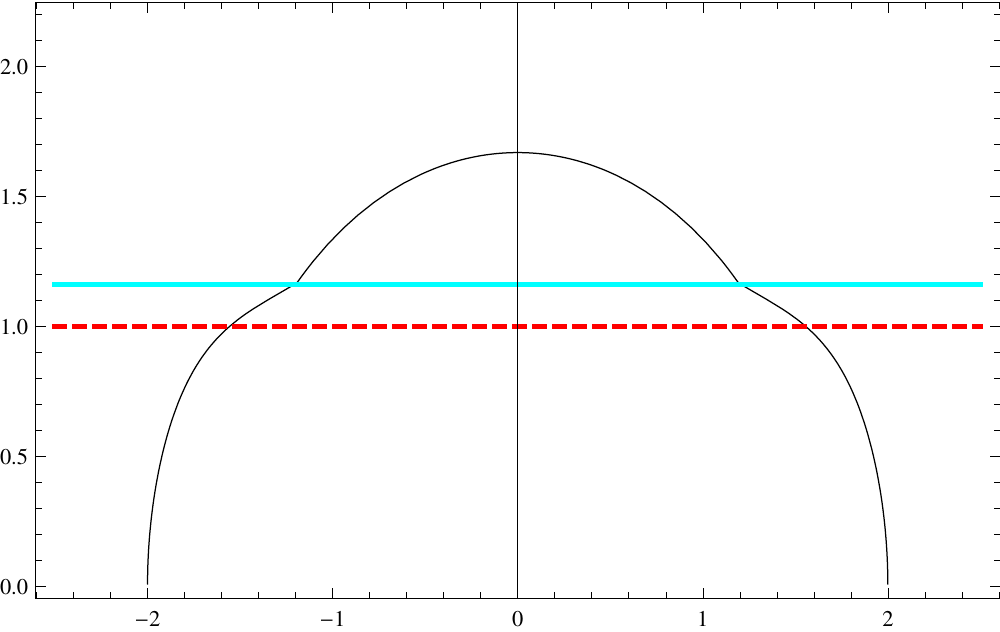}}\qquad
    \subfigure[$b=0.4$ at $t=1.54$]{\includegraphics[height=1.1in]{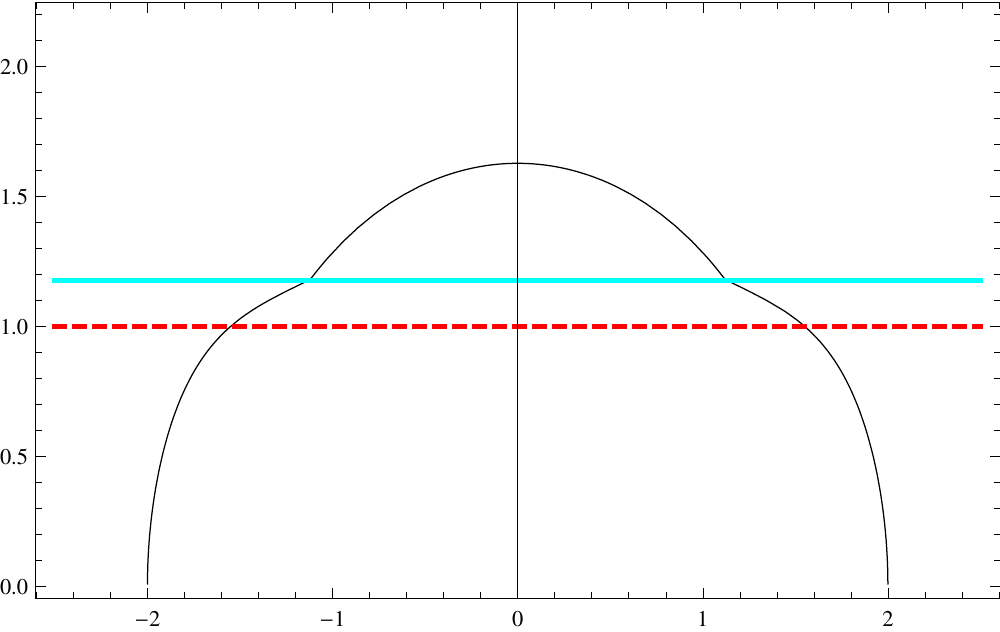}}\qquad
    \subfigure[$b=1$ at $t=1.54$]{\includegraphics[height=1.1in]{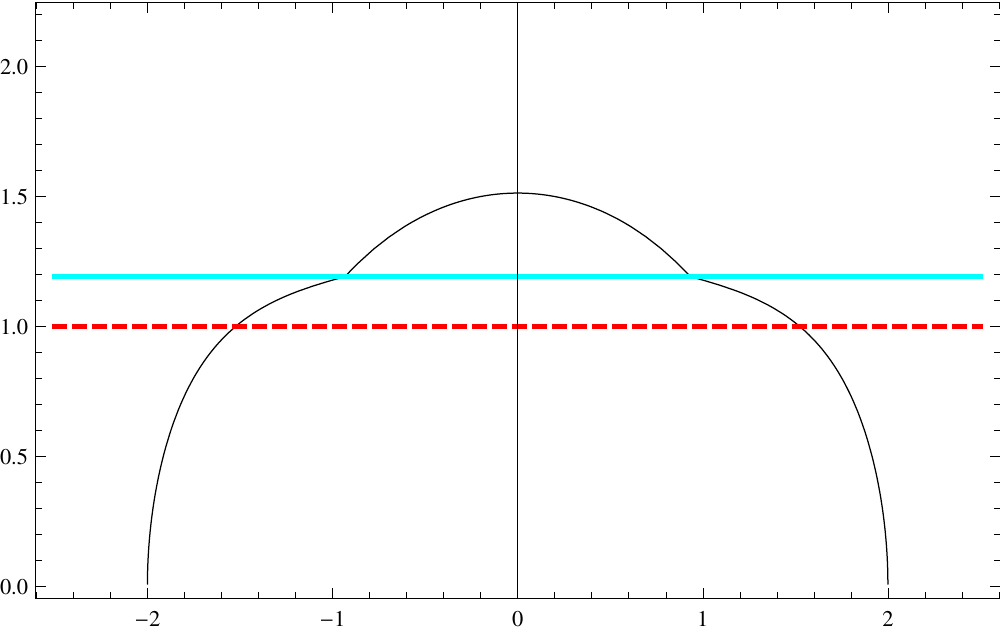}}\\
    \subfigure[$b=0$ at $t=1.82$]{\includegraphics[height=1.1in]{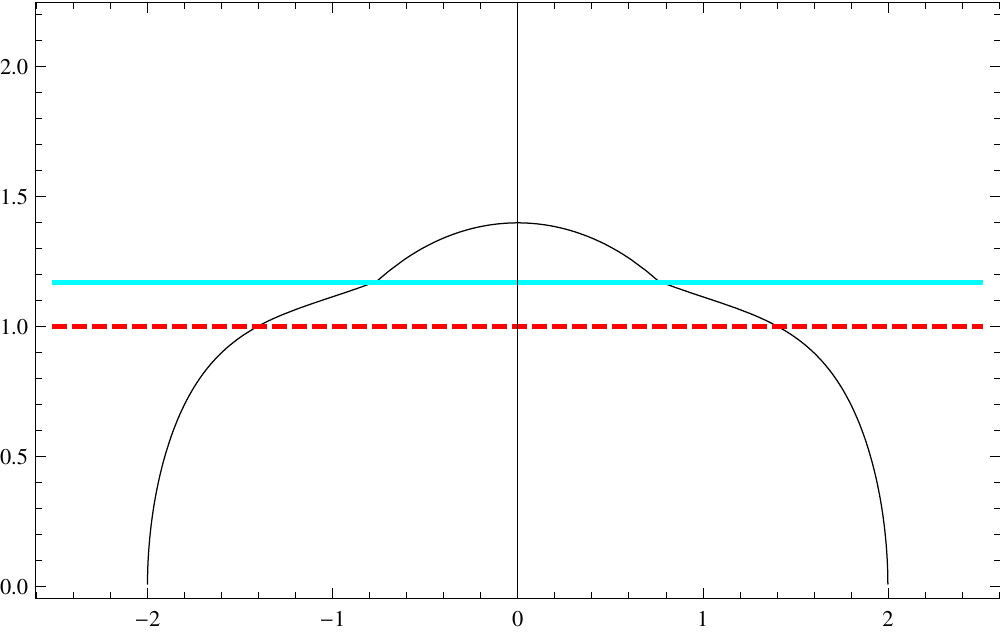}}\qquad
    \subfigure[$b=0.4$ at $t=1.82$]{\includegraphics[height=1.1in]{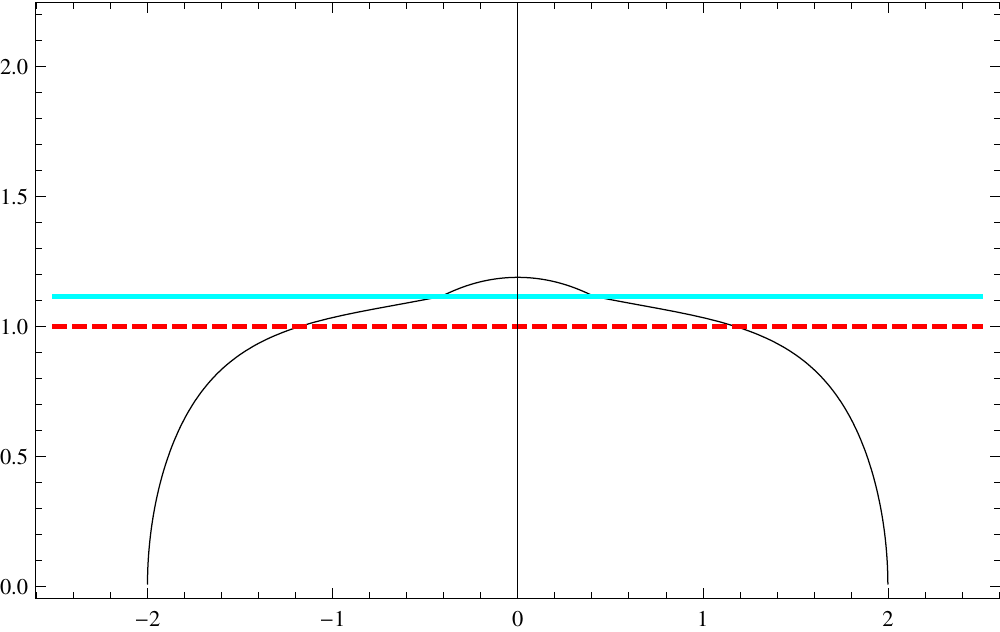}}\qquad
    \subfigure[$b=1$ at $t=1.82$]{\includegraphics[height=1.1in]{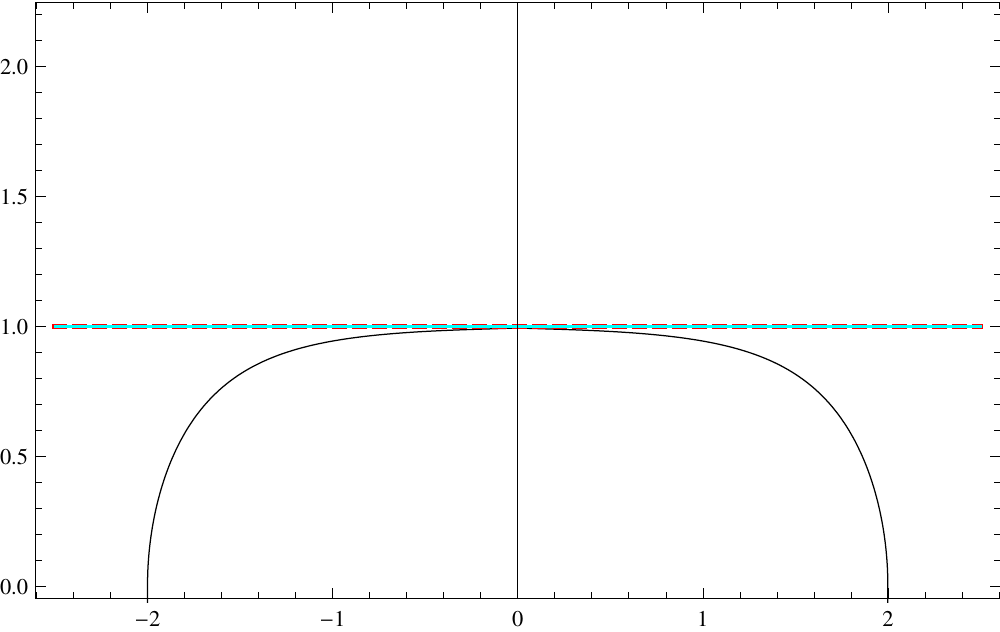}}\\
    \caption{A sequence of snapshots of the time evolution of geodesic profiles and the shell of charged dust described by the Vaidya-BI-AdS metric to form a black brane at $z_h=1$ at late times, for different values of the inverse BI parameter $b$ and fixed charge $Q=1$. In all cases the separation of the boundary points is $\ell=4$. Each column, from top to bottom, indicates the time evolution for a given value of $b$. The cyan line indicates the position of the shell in each case, while the dashed red line is an imaginary line denoting the position of the (still to be formed) black brane horizon.}\label{fig:geoprofile}
\end{figure}

\begin{figure}[htbp]% Thermalization at Q fixed
    \centering
    \subfigure[$Q=0.5$]{\includegraphics[height=1.24in]{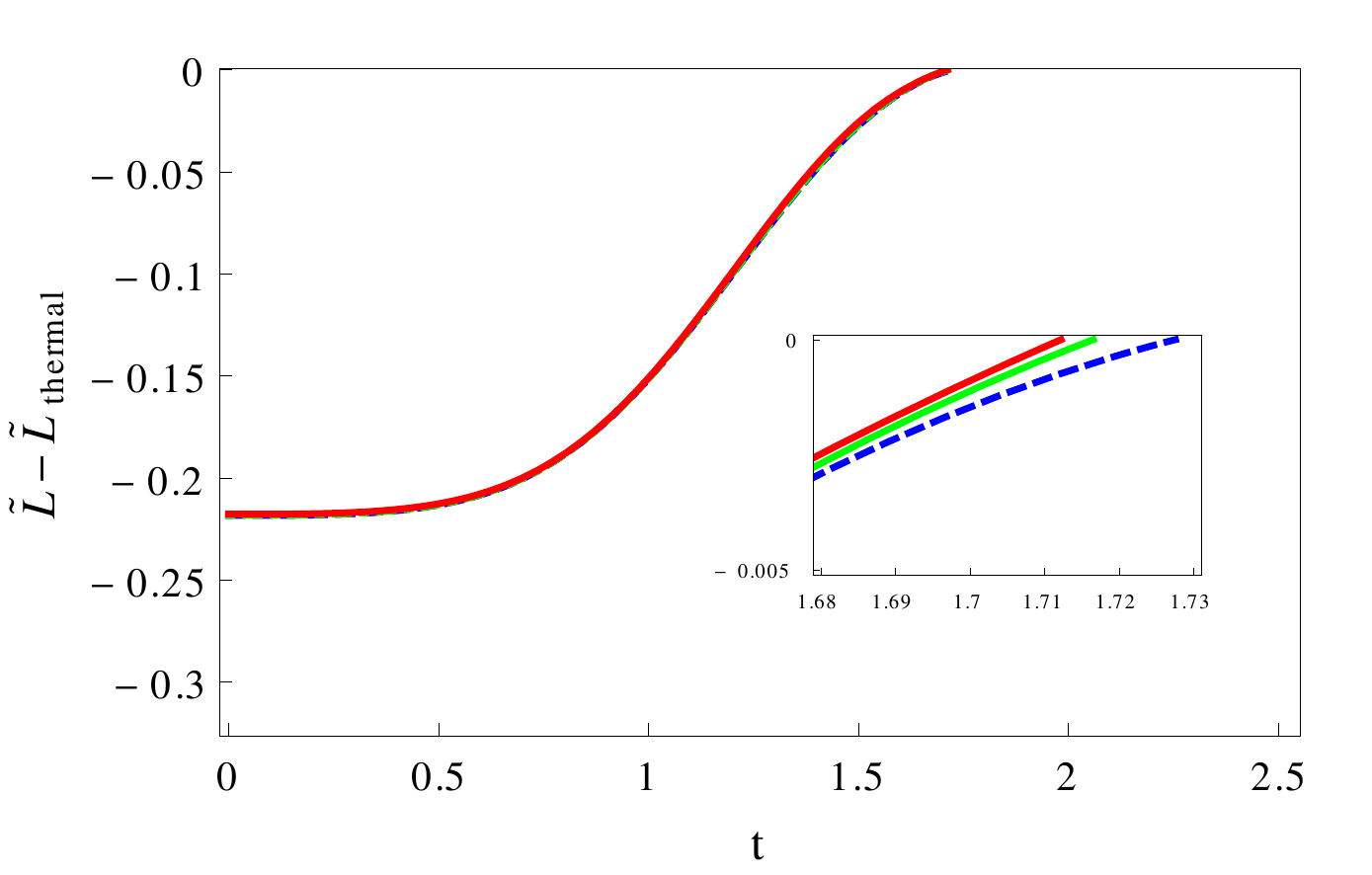}}
    \subfigure[$Q=1$]{\includegraphics[height=1.24in]{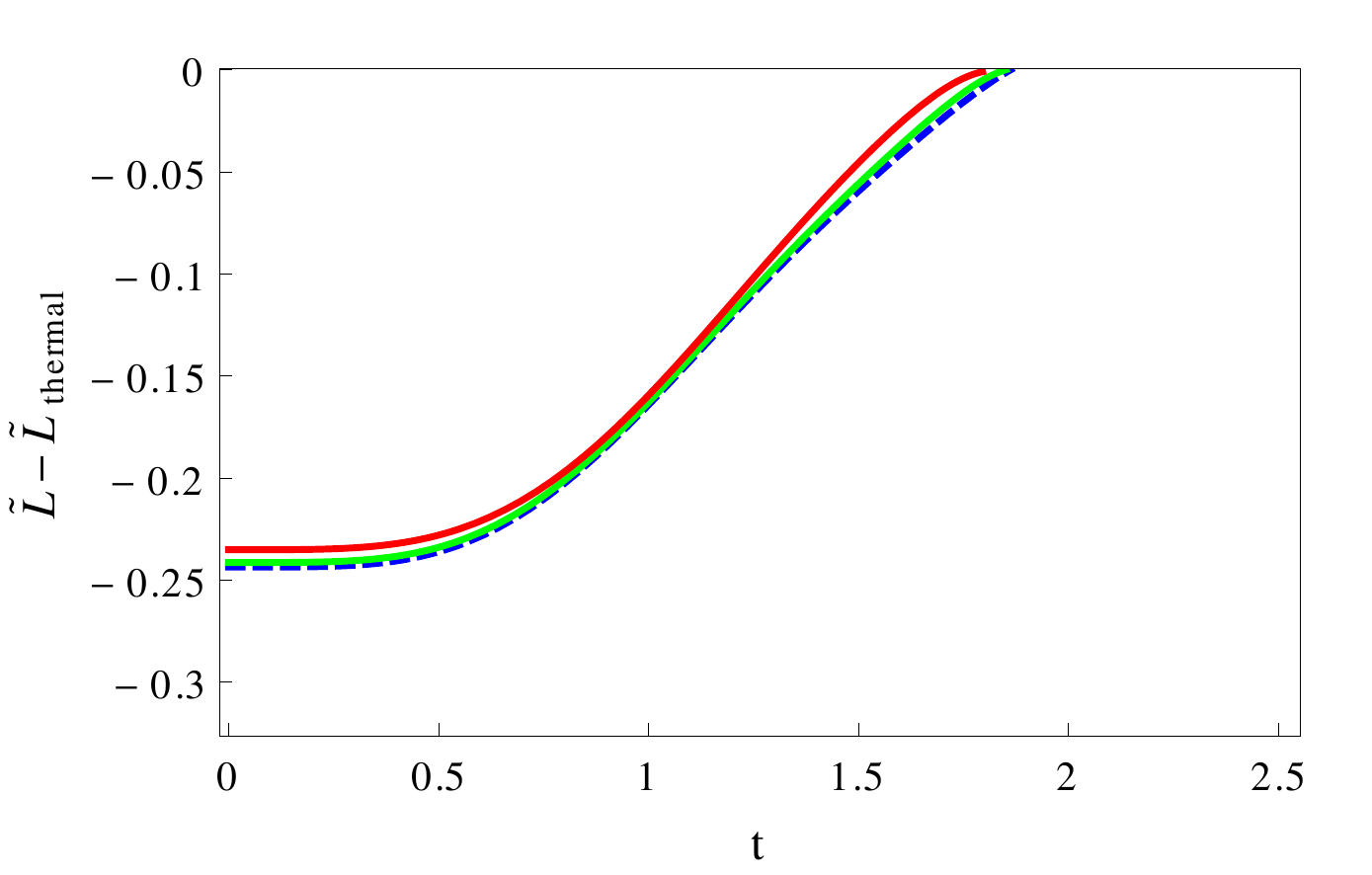}}
    \subfigure[$Q=\sqrt{2}$]{\includegraphics[height=1.24in]{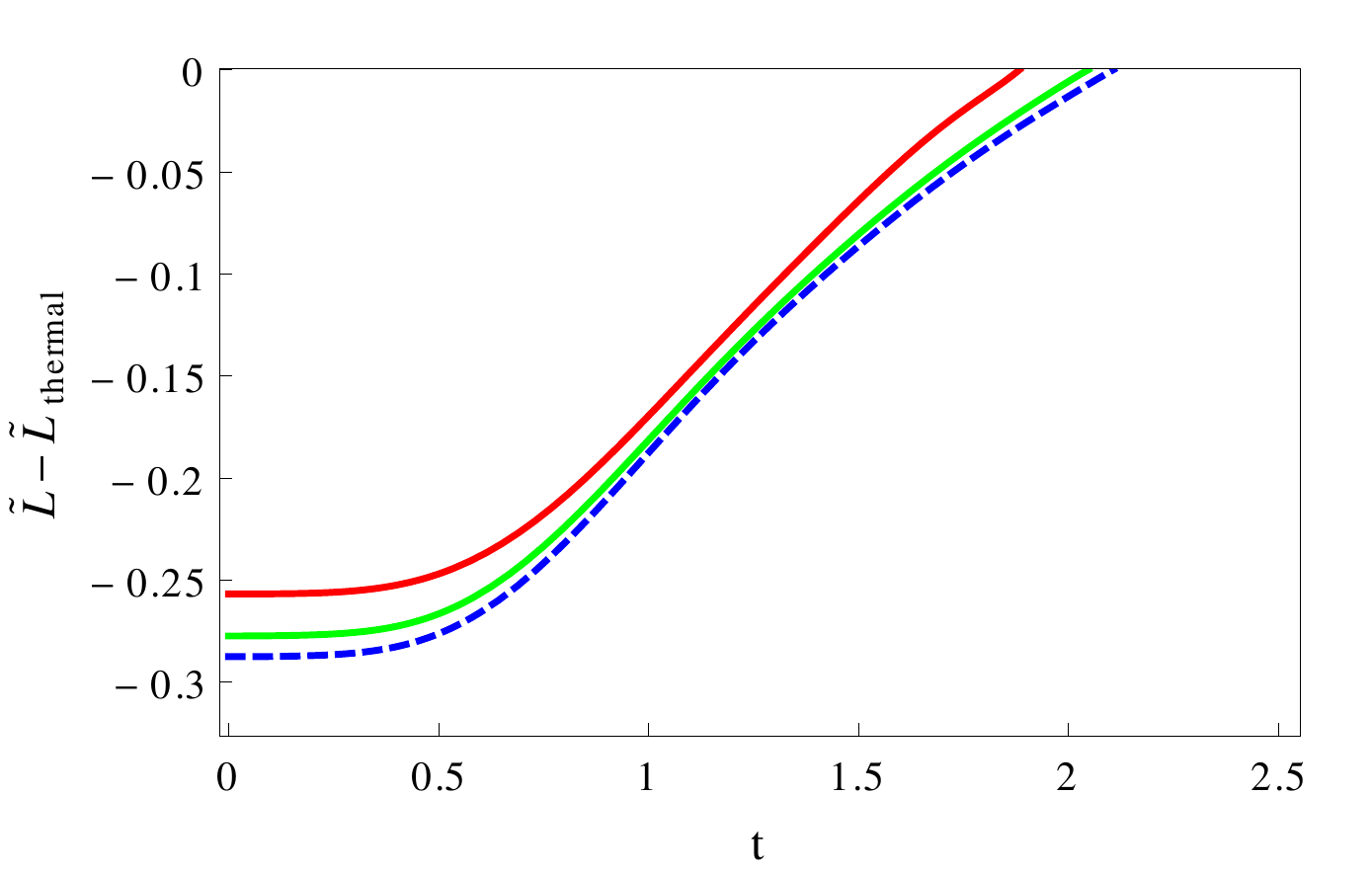}}\\
    \caption{Thermalization curves of the renormalized geodesic lengths in the Vaidya-BI-AdS spacetime at fixed charges $Q$ for different inverse BI parameters $b$. Here $b=0$ (dashed blue) is always the bottom curve, $b=0.4$ (green) is the middle curve, and $b=1$ (red) is the top one. The spatial separation of the boundary points is $\ell=4$.}\label{fig:geoQfixed}
\end{figure}

Before showing the thermalization curves, in Figure \ref{fig:geoprofile} we present an intuitive simple view of the effect of the BI parameter on the thermalization. It consists of a sequence of snapshots of the time evolution of geodesic profiles as well as the shell of charged dust described by the Vaidya-BI-AdS metric to form a black brane at $z_h=1$ at late times, for different values of the inverse BI parameter $b$ and fixed charge $Q=1$. Each column, top to bottom, corresponds to the time evolution for a given value of $b$. It is found that at the early stages of the evolution, up to $t\sim1$, the value of $b$ has little effect on the dynamics. After that, $b$ plays a crucial role in the evolution.
We see that the bigger $b$ is (i.e., columns to the right), the sooner the black hole is formed. This is clear from the bottom line of the picture, corresponding to boundary time $t=1.82$, where the $b=1$ black brane has just formed while the $b=0.4$ one is about to form and the $b=0$ one still needs some time to do so. This is a hint that the thermalization of the dual boundary field theory occurs sooner as $b$ increases, what indeed will be confirmed in Figure \ref{fig:geoQfixed}.

\begin{figure}[htbp]% Thermalization at b fixed
    \centering
    \subfigure[$b=0$]{\includegraphics[height=1.24in]{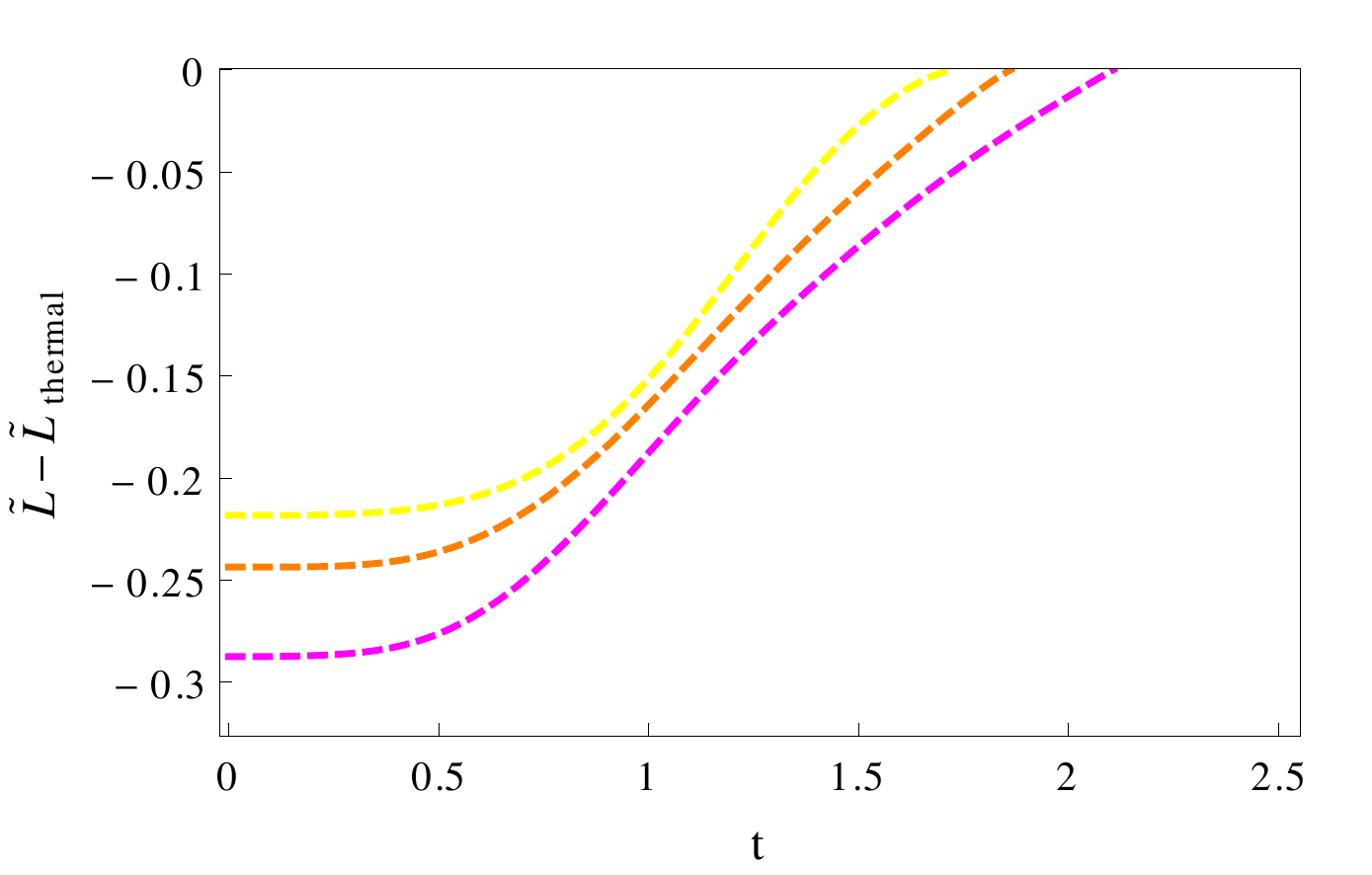}}
    \subfigure[$b=0.4$]{\includegraphics[height=1.24in]{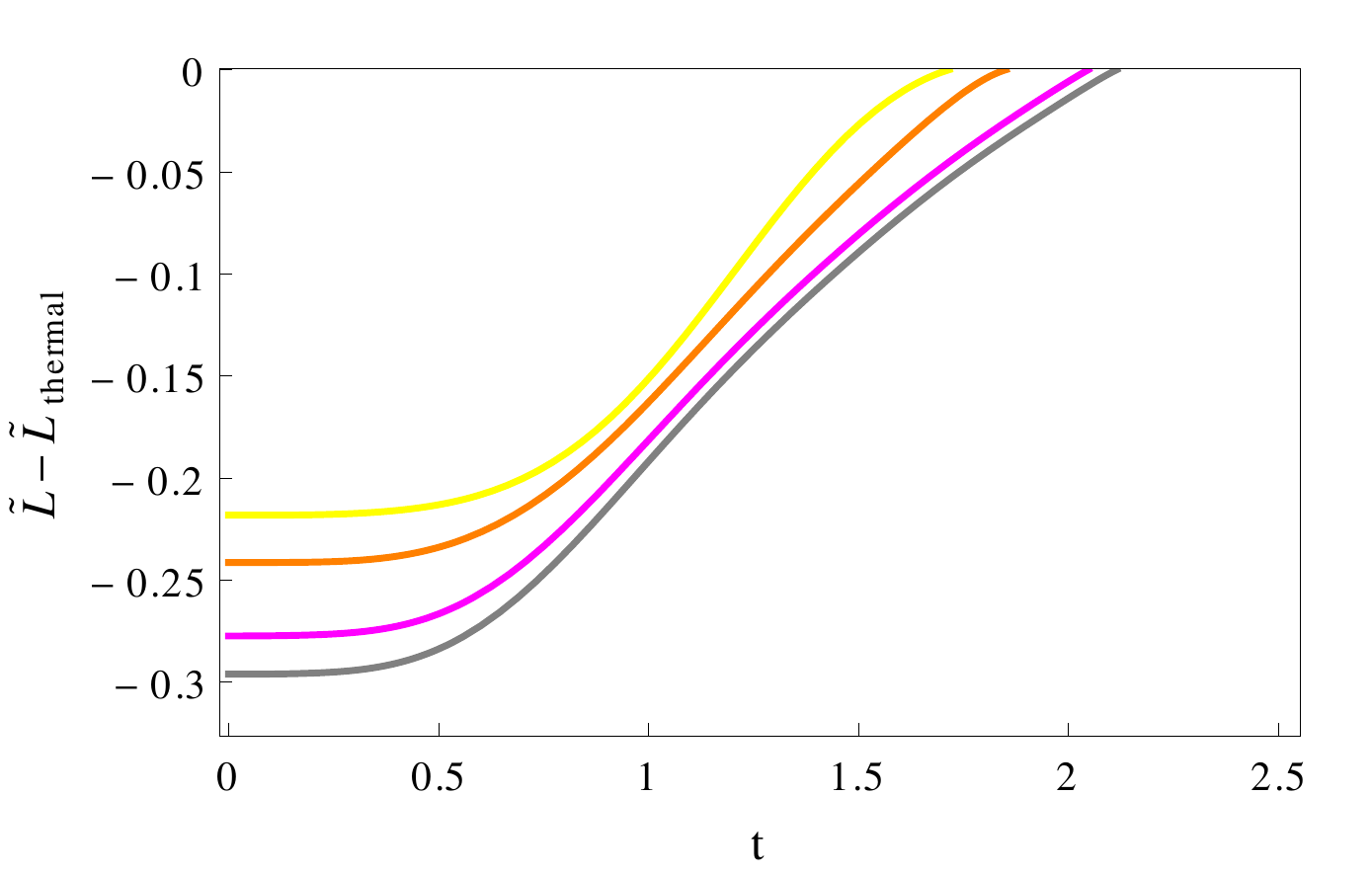}}
    \subfigure[$b=1$]{\includegraphics[height=1.24in]{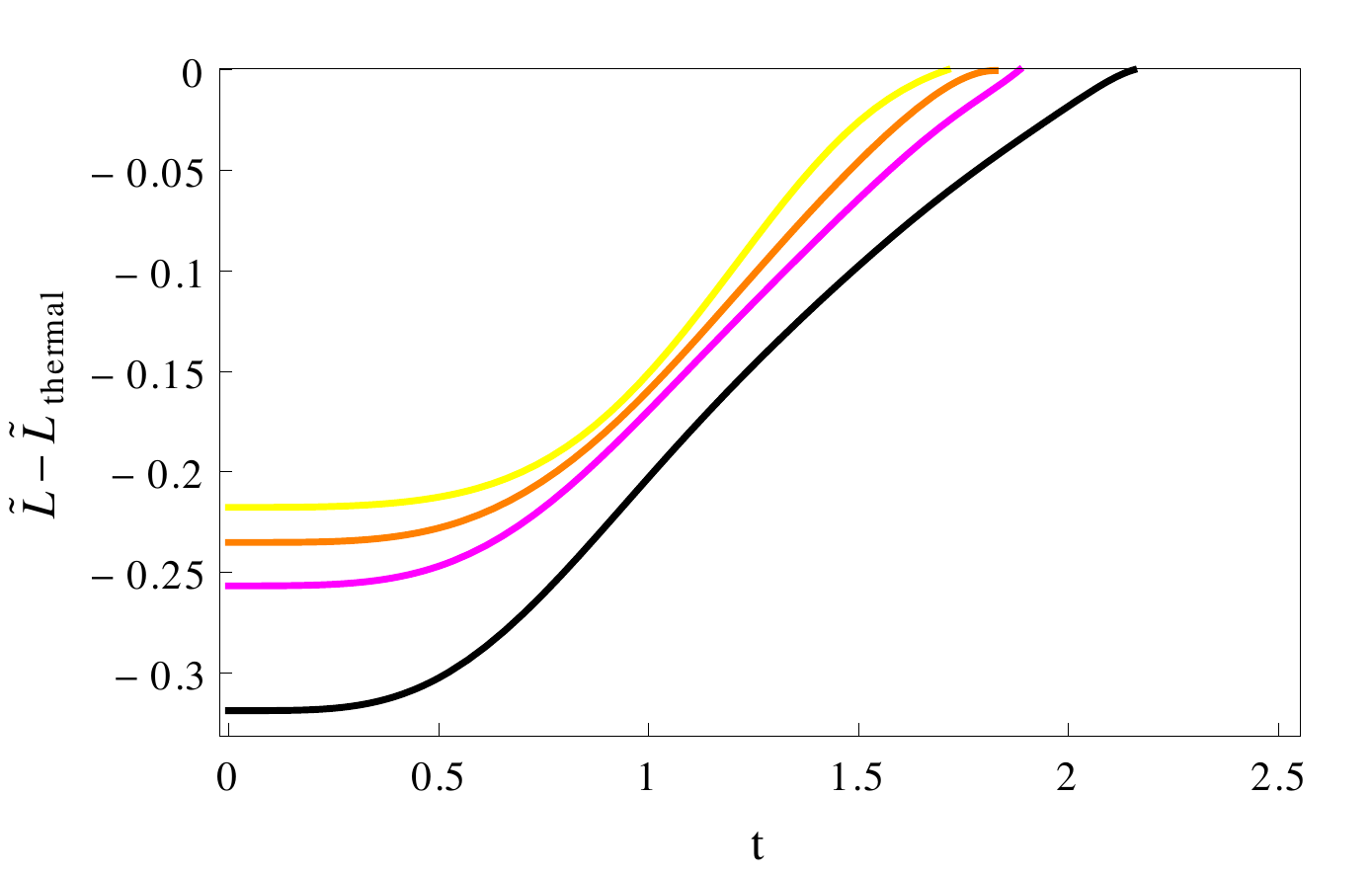}}\\
    \caption{Thermalization curves of the renormalized geodesic lengths in the Vaidya-BI-AdS spacetime at fixed values of $b$ for different charges $Q$, from $Q=0.5$ to the corresponding extremal value $Q_{ext}(b)=\sqrt{2+3b^2}$. Figure (a) shows the Maxwell limiting case, with $Q=0.5$ (yellow) in the top, $Q=1$ (orange) in the middle and $Q=\sqrt{2}$ (magenta) in the bottom. In (b) and (c) we have the same values of charge together with the extremal values $Q=\sqrt{2.48}$ (gray) and $Q=\sqrt{5}$ (black), respectively. The spatial separation of the boundary points is $\ell=4$ for all the cases.}\label{fig:geobfixed}
\end{figure}

In Figures \ref{fig:geobfixed} and \ref{fig:geoQfixed} we show the thermalization curves for the renormalized geodesic lengths with varying charge at fixed $b$ and varying $b$ at fixed charges, respectively. Instead of plotting point by point all the results obtained as described above, we find more instructive to fit those points using some polynomial function and plot the resulting curve. Details about the fits will be given below. We use dashed curves in all the plots hereinafter to highlight the Maxwell limiting case studied in \cite{galante}. The thermalized (final) state corresponding to the completely formed black brane is reached in each case when the curve touches the zero point of the vertical axis. 
The effect of the charge $Q$ on the thermalization is clear from Figure \ref{fig:geobfixed}. As $Q$ grows, the thermalization time increases, meaning that the dual field theory thermalizes later. This had already been pointed out in \cite{galante} for the case of Maxwell electrodynamics ($b=0$), and now we show that the same holds for BI nonlinear electrodynamics. Since the charge corresponds to the chemical potential, this means that the smaller the chemical potential is, the faster is the pair production and the screening effect takes over in an easier way. This is compatible with lower dimensional models, where screening effects are known to prevail over confinement \cite{elcio1,elcio2}.
The second, more interesting result, is the effect of the inverse BI parameter $b$ shown in Figure \ref{fig:geoQfixed}. As one can see, increasing $b$ decreases the thermalization time, which means that the more nonlinear the bulk theory is, the sooner its dual field theory thermalizes. This confirms our intuition coming from the analysis of the geodesic profiles and shell motion in Figure \ref{fig:geoprofile}. Such a behavior is similar to the effect of the Gauss-Bonnet parameter on the thermalization reported in \cite{chinesesGB}. As we discuss in the conclusions, this seems to be a general feature of introducing extra derivatives in the bulk theory. The numerical values obtained for the thermalization times are summarized in Table \ref{tab:geotimes}.

\begin{table}[htbp]
\centering
\begin{tabular}{c|c|c|c|c}
\cline{2-4}
                            & $b=0$   & $b=0.4$ & $b=1$   &  \\ \cline{1-4}
\multicolumn{1}{|c|}{$Q=0.5$} & $1.728$ & $1.717$ & $1.712$ &  \\ \cline{1-4}
\multicolumn{1}{|c|}{$Q=1$}   & $1.863$ & $1.852$ & $1.796$ &  \\ \cline{1-4}
\multicolumn{1}{|c|}{$Q=\sqrt{2}$}   & $2.108$ & $2.048$ & $1.887$ &  \\ \cline{1-4}
\multicolumn{1}{|c|}{$Q=\sqrt{2.48}$}   & -- & $2.116$ & -- &  \\ \cline{1-4}
\multicolumn{1}{|c|}{$Q=\sqrt{5}$}   & -- & -- & $2.156$ &  \\ \cline{1-4}
\end{tabular}
\caption{Summary of the numerical values obtained for the thermalization times of the renormalized geodesic length curves shown in Figures \ref{fig:geobfixed} and \ref{fig:geoQfixed}.}\label{tab:geotimes}
\end{table}

\begin{figure}[htbp]% Thermalization velocities
    \centering
    \subfigure[$Q=1$]{\includegraphics[height=1.35in]{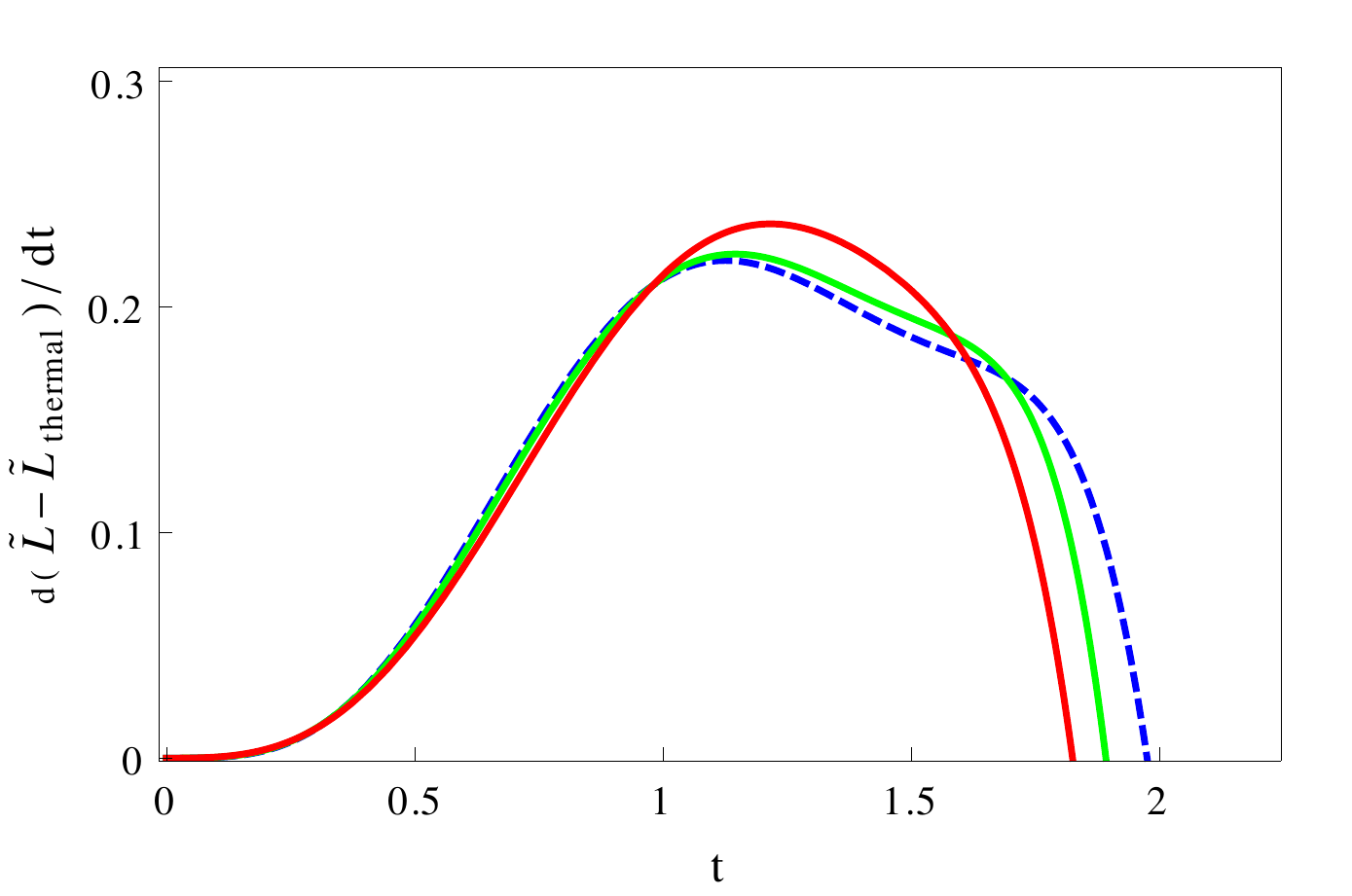}}\label{fig:veloell4Q1}\qquad
    \subfigure[$b=0.4$]{\includegraphics[height=1.35in]{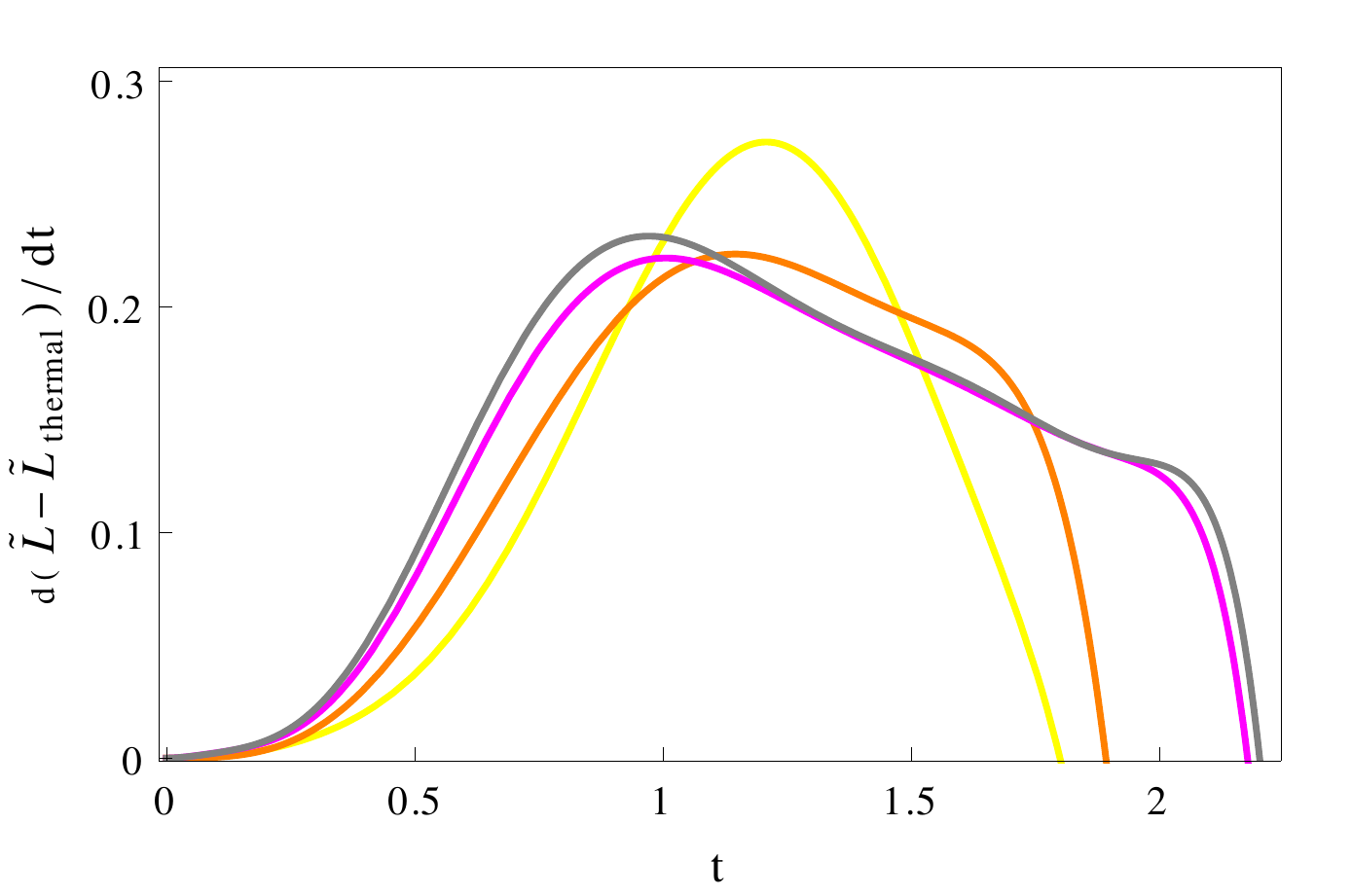}}\label{fig:veloell4b04}
    \caption{Thermalization velocities of the renormalized geodesic lengths at both fixed charge and fixed $b$. The dashed blue, green, and red curves in (a) correspond respectively to $b=0,0.4,1$. The curves in (b) correspond to $Q=0.5$ (yellow), $Q=1$ (orange), $Q=\sqrt{2}$ (magenta) and $Q=\sqrt{2.48}$ (gray). The spatial separation of the boundary points is $\ell=4$.}\label{fig:geovelo}
\end{figure}

Having smooth fit functions for all the sets of numerical data we can use them to study the thermalization velocities $\frac{d}{dt}(\tilde{\mathcal{L}}-\tilde{\mathcal{L}}_{\textrm{thermal}})$ aiming for more details of the nonequilibrium process. These are plotted in Figure \ref{fig:geovelo} (only for the cases $Q=1$ fixed and $b=0.4$ fixed, respectively, to avoid unnecessary repetitions). We notice from the velocity curves the existence of a phase transition point at the middle stage of the thermalization, which divides the process into an accelerating and a decelerating phase. Furthermore, we see that the phase transition point is shifted depending on the values of $b$ and $Q$. Figure (\ref{fig:veloell4Q1}) shows that increasing the value of $b$ causes a delay in the phase transition point, meaning that the accelerated phase lasts longer for the $b=1$ theory. This is to be contrasted with the fact that the $b=1$ theory is the first to thermalize, indicating that the dynamical process in this case consists of a slowly accelerating phase followed by a quick deceleration towards the equilibrium state. On the other hand, Figure (\ref{fig:veloell4b04}) shows that the charge has the opposite effect, i.e., as $Q$ increases the phase transition point arrives earlier. In other words, for large values of $Q$ (or $\mu/T$ in the boundary field theory) the thermalization process consists of a quick accelerating phase followed by a slowly decelerating phase to the final state.

\begin{figure}[htbp]% Comparation data/fits
    \centering
    \subfigure[$Q=1,b=1$]{\includegraphics[height=1.35in]{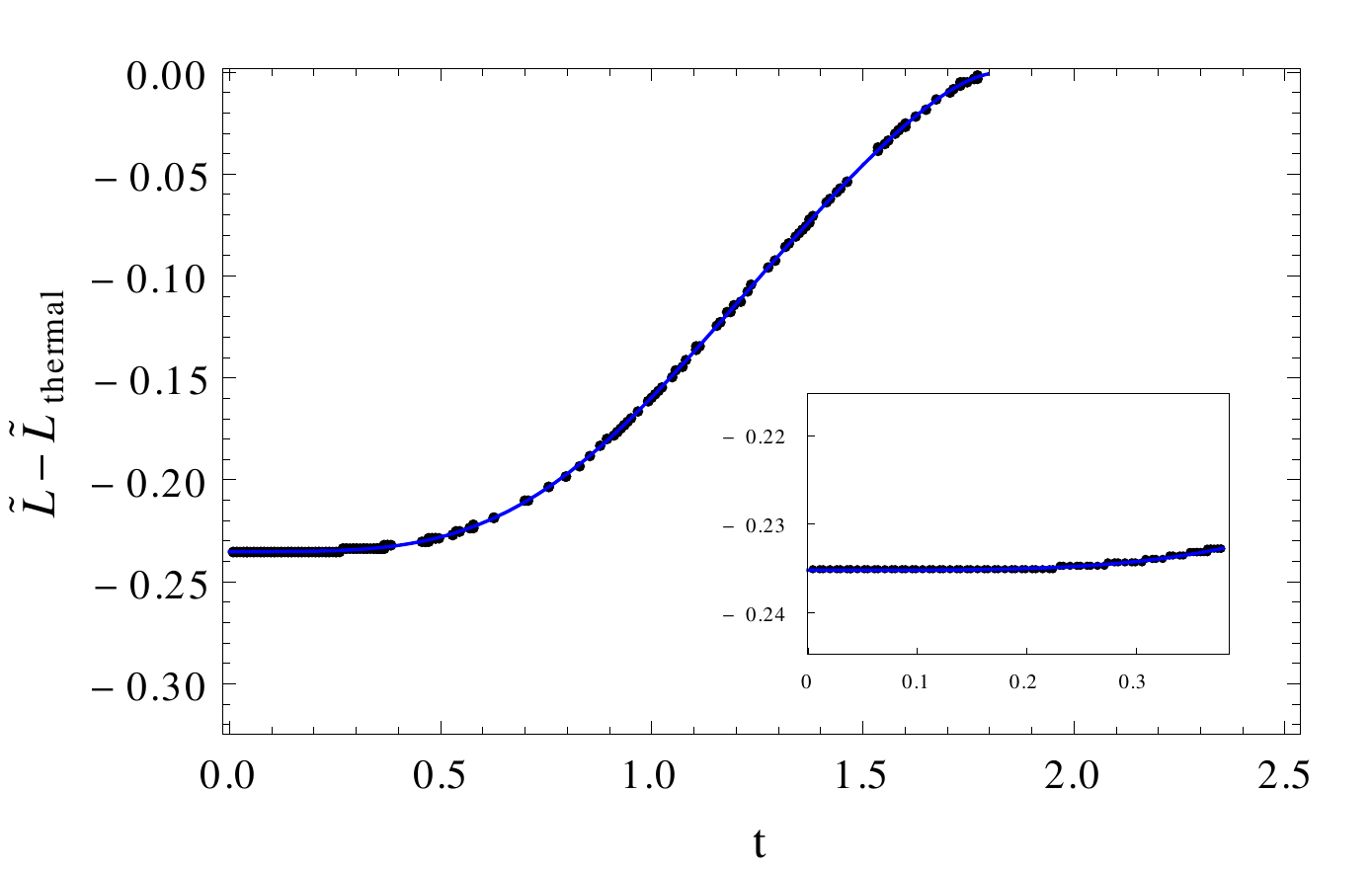}}\qquad
    \subfigure[$Q=\sqrt{2},b=0.4$]{\includegraphics[height=1.35in]{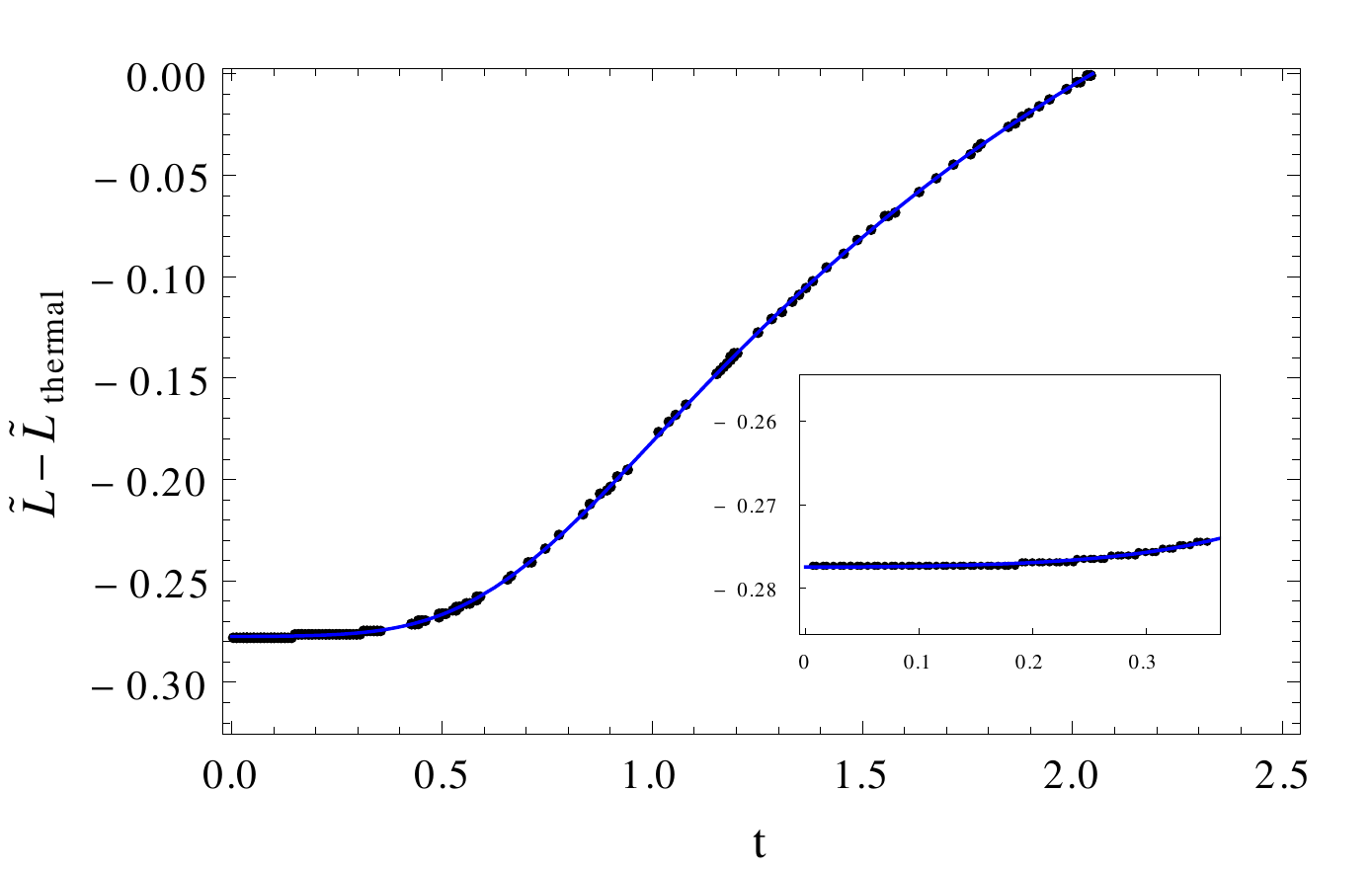}}
    \caption{A comparison between the numerical results and the polynomial fits for the geodesic lengths. The inset emphasizes the constant behavior of the curve at initial times.}\label{fig:geofits}
\end{figure}

It should be stressed that in this work, in contrast to the authors of \cite{chinesesGBQ}, we find no evidence for a negative thermalization velocity at initial times. They argue that the velocity should be negative in the very beginning of the evolution corresponding to a\lq\lq quantum\rq\rq stage of the nonequilibrium process, which soon becomes\lq\lq classical\rq\rq once the velocity becomes positive. In our case (Figure \ref{fig:geovelo}) all the thermalization velocities start from zero and increase monotonically until the phase transition point, indicating that nothing particularly odd seems to happen at the initial stages of the thermalization process. This is also clear from the comparison of the numerical results with the fitting functions presented in Figure \ref{fig:geofits}. The zoomed region shows that the numerical points sit all over a horizontal line for initial times (up to $\sim0.2$) and therefore there is no reason for a non-vanishing slope at such stage. For that reason, we use for our fit functions degree $9$ polynomials $f(t)=\sum_{n=0}^{9}\alpha_nt^n$ with the first powers of $t$ ($\alpha_1,\alpha_2$) set to zero in order to ensure the strictly constant behavior $f(t)\sim \alpha_0$ up to $t\sim0.2$. This allows us to make an accurate fit of the whole set of numerical points, which after all are the ones carrying the physical information. 

\subsection{Renormalized minimal area surfaces}

In this section, we numerically solve the equations of motion (\ref{eq:areaeqs}) for the minimal area surfaces in order to track their time evolution. The strategy will follow closely that of the renormalized geodesic lengths done in the previous subsection, so we will not repeat all the details on the numerical procedure as well as the fixing of free parameters since they are essentially identical.

Again we choose the test values $b=0,0.4,1$ and $Q=0.5,\ldots,Q_{ext}(b)$ for the numerical analysis. The procedure is the same as before, i.e., for a given $(b,Q)$ pair we solve the geodesic equations (\ref{eq:areaeqs}) subject to the boundary conditions (\ref{eq:bcmod}) characterized by a pair of values $(v_*,z_*)$. We repeat this for various pairs of $(v_*,z_*)$ and collect just those yielding a Wilson loop with side $\ell=2$ (keep in mind that the other side $R$ does not influence in our analysis).\footnote{Remember that $\ell$ is determined from the numerics, via equation (\ref{eq:bc}), so we again use he same criterion of $\pm0.0005$ for what we mean by \lq\lq$\ell=2$\rq\rq.} Each of the collected solutions will correspond to a different stage of the time evolution determined by calculating the boundary time $t$ via equation (\ref{eq:bc}). Then we integrate each of the collected solutions using equation (\ref{eq:nambuonshell}), subtract the universal divergent part to obtain $\mathcal{A}_{ren}$, and construct a list of points $(t,\tilde{\mathcal{A}}-\tilde{\mathcal{A}}_{\textrm{thermal}})$ to be plotted against $t$. Here $\tilde{\mathcal{A}}\equiv\mathcal{A}_{ren}/(R\ell\pi^{-1})$ is a dimensionless quantity independent of the dimensions of the boundary Wilson loop and $\tilde{\mathcal{A}}_{\textrm{thermal}}$ is the corresponding thermal value.

\begin{figure}[thbp]% Area profiles
    \centering
    \subfigure[$b=0$ at $t=0.45$]{\includegraphics[height=1.4in]{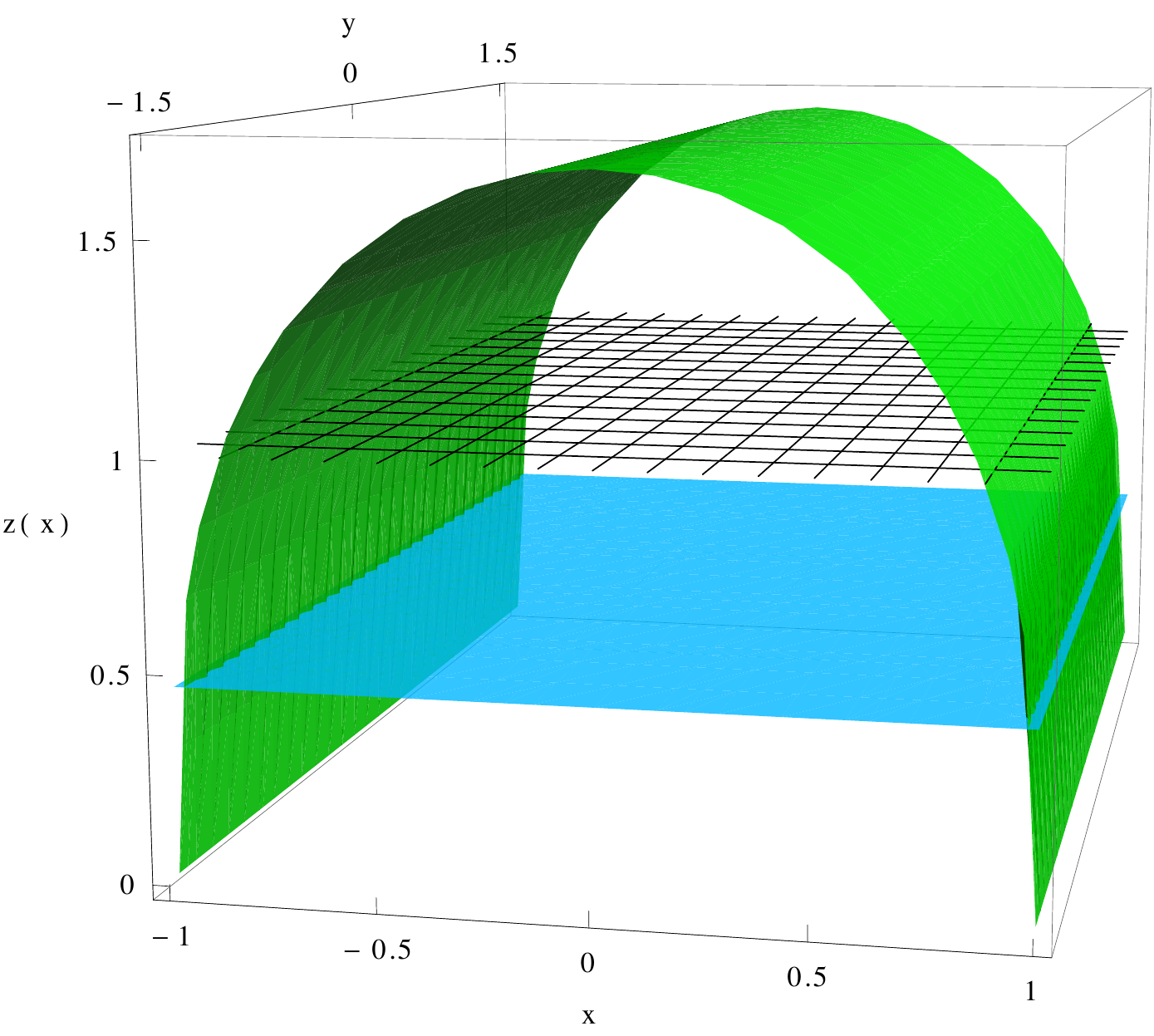}}\qquad
    \subfigure[$b=0.4$ at $t=0.45$]{\includegraphics[height=1.4in]{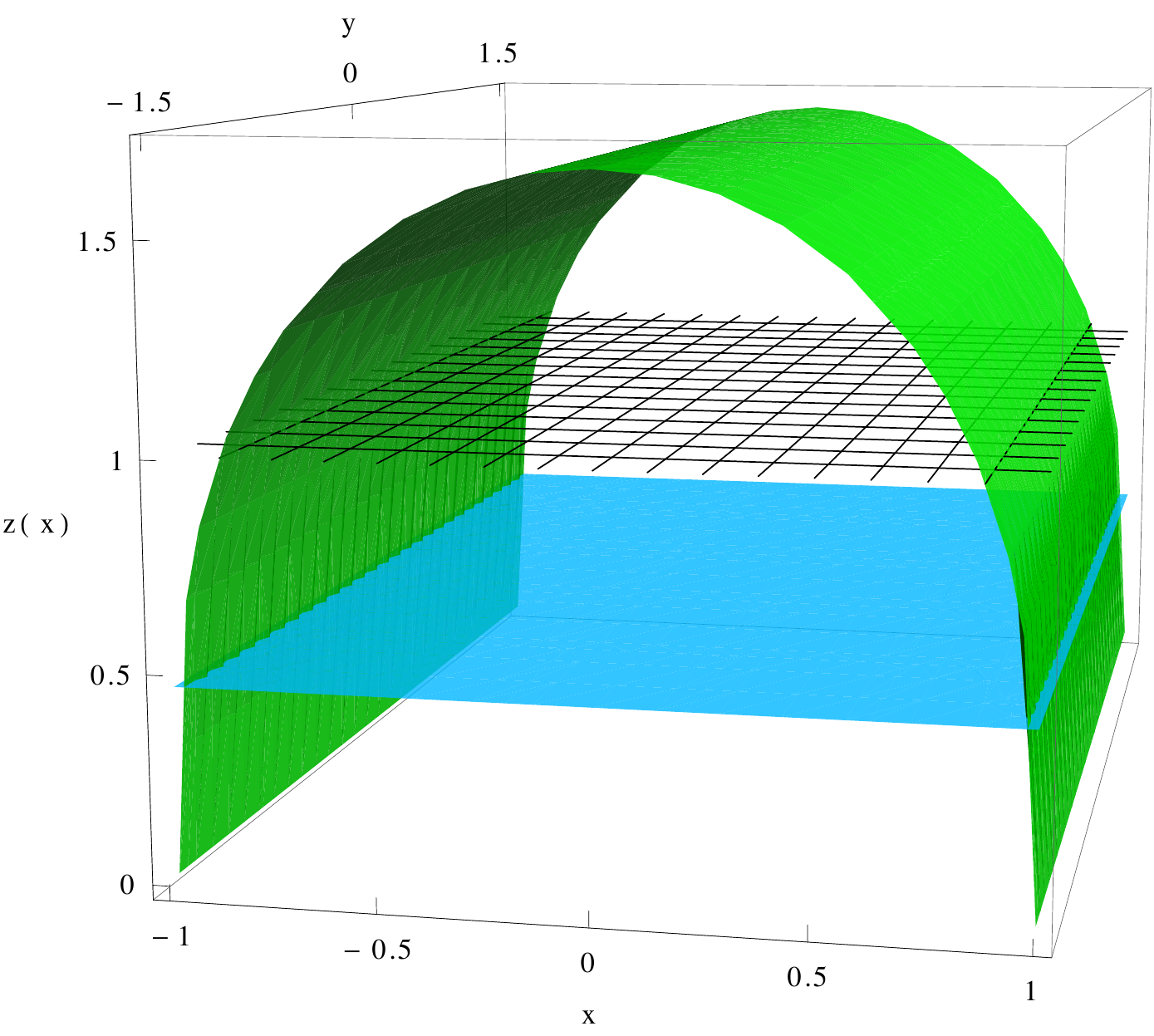}}\qquad
    \subfigure[$b=1$ at $t=0.45$]{\includegraphics[height=1.4in]{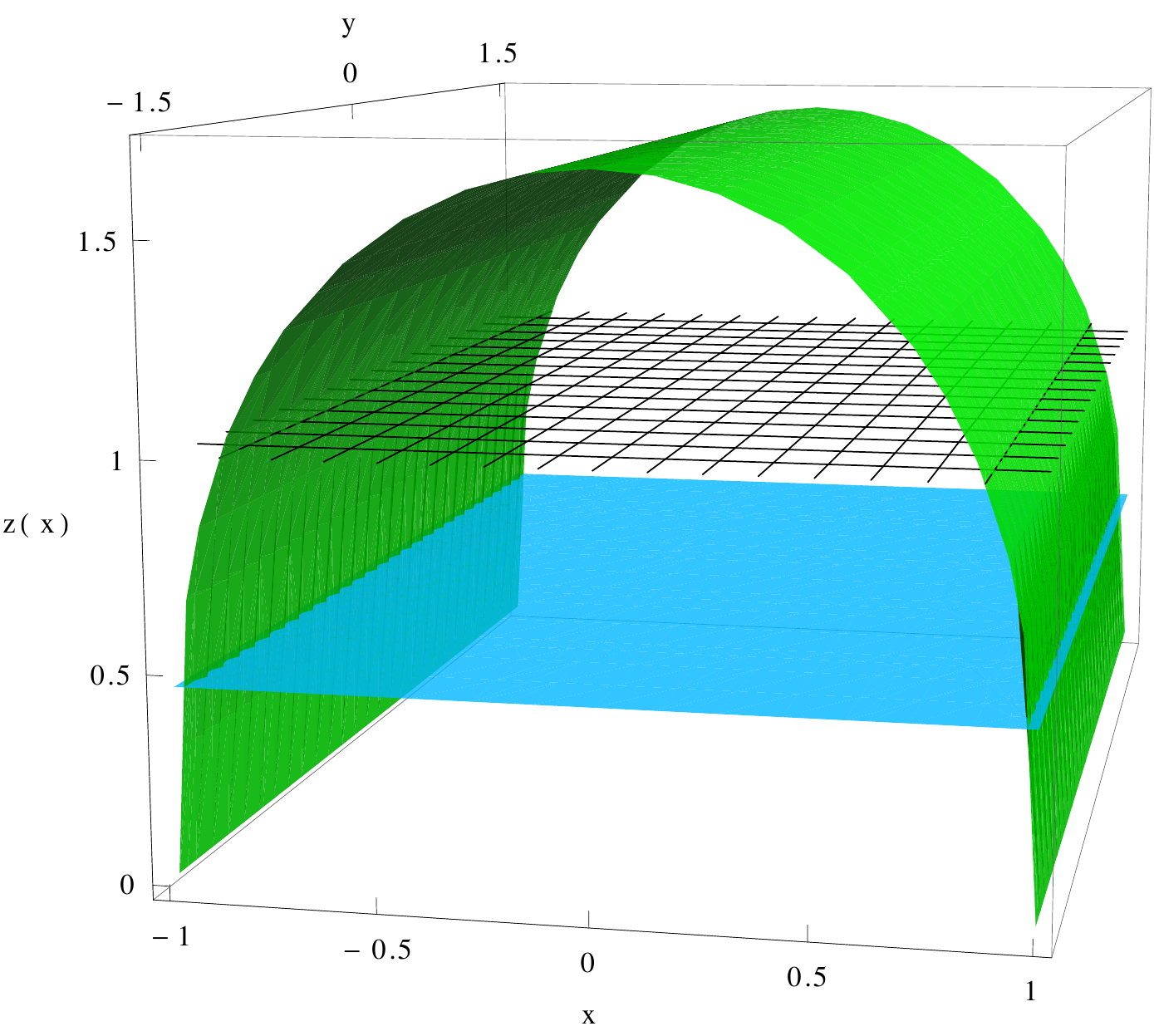}}\\
    \subfigure[$b=0$ at $t=0.96$]{\includegraphics[height=1.4in]{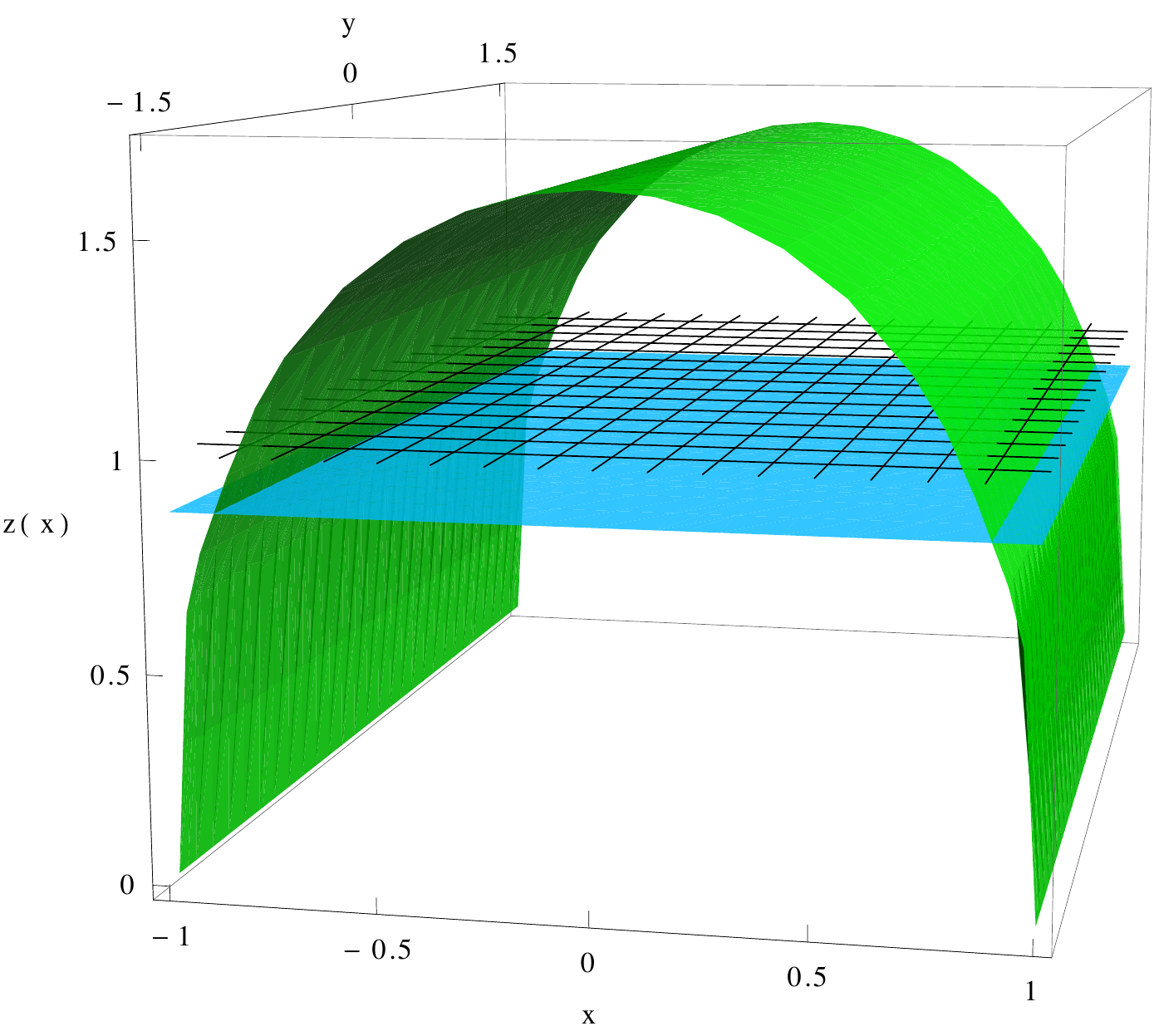}}\qquad
    \subfigure[$b=0.4$ at $t=0.96$]{\includegraphics[height=1.4in]{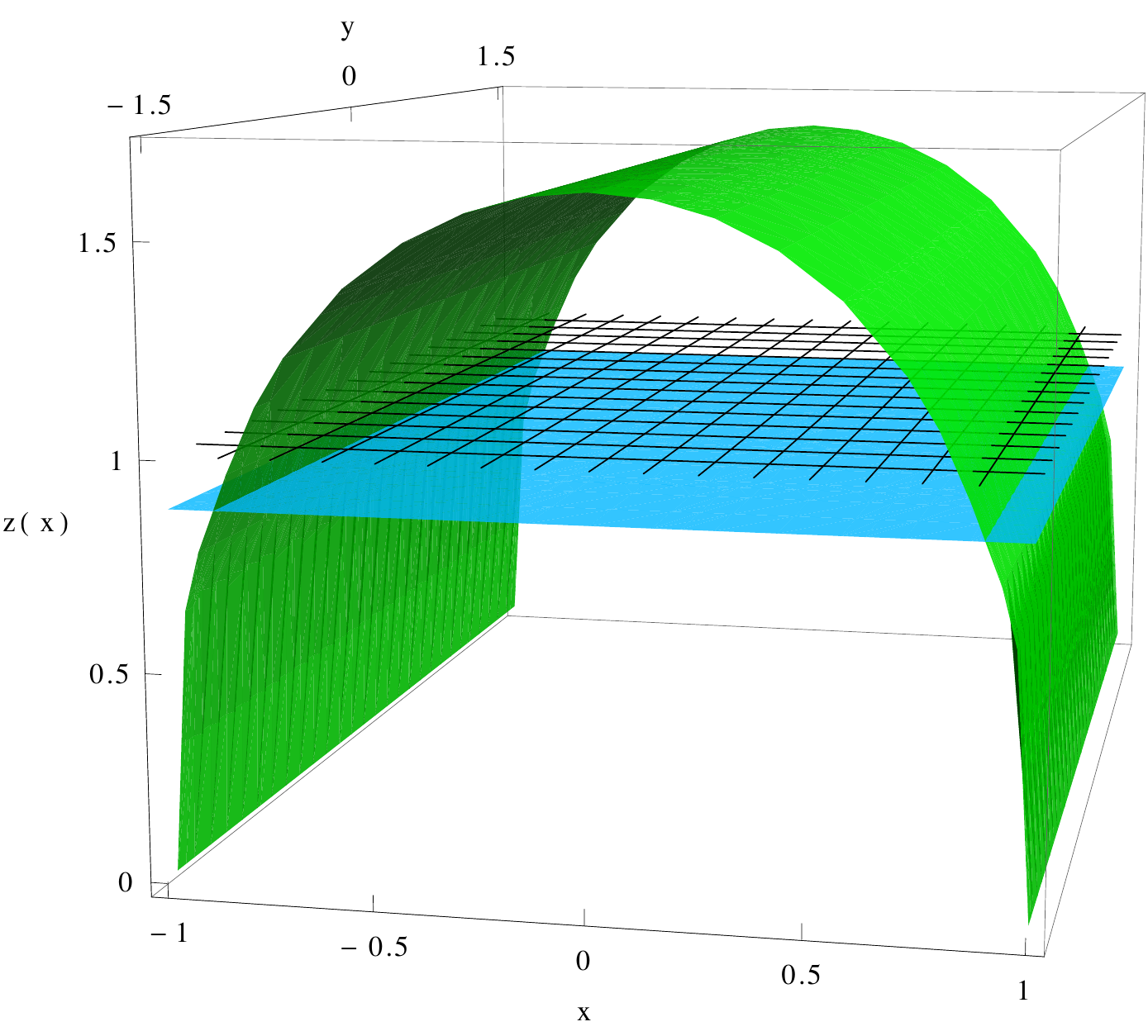}}\qquad
    \subfigure[$b=1$ at $t=0.96$]{\includegraphics[height=1.4in]{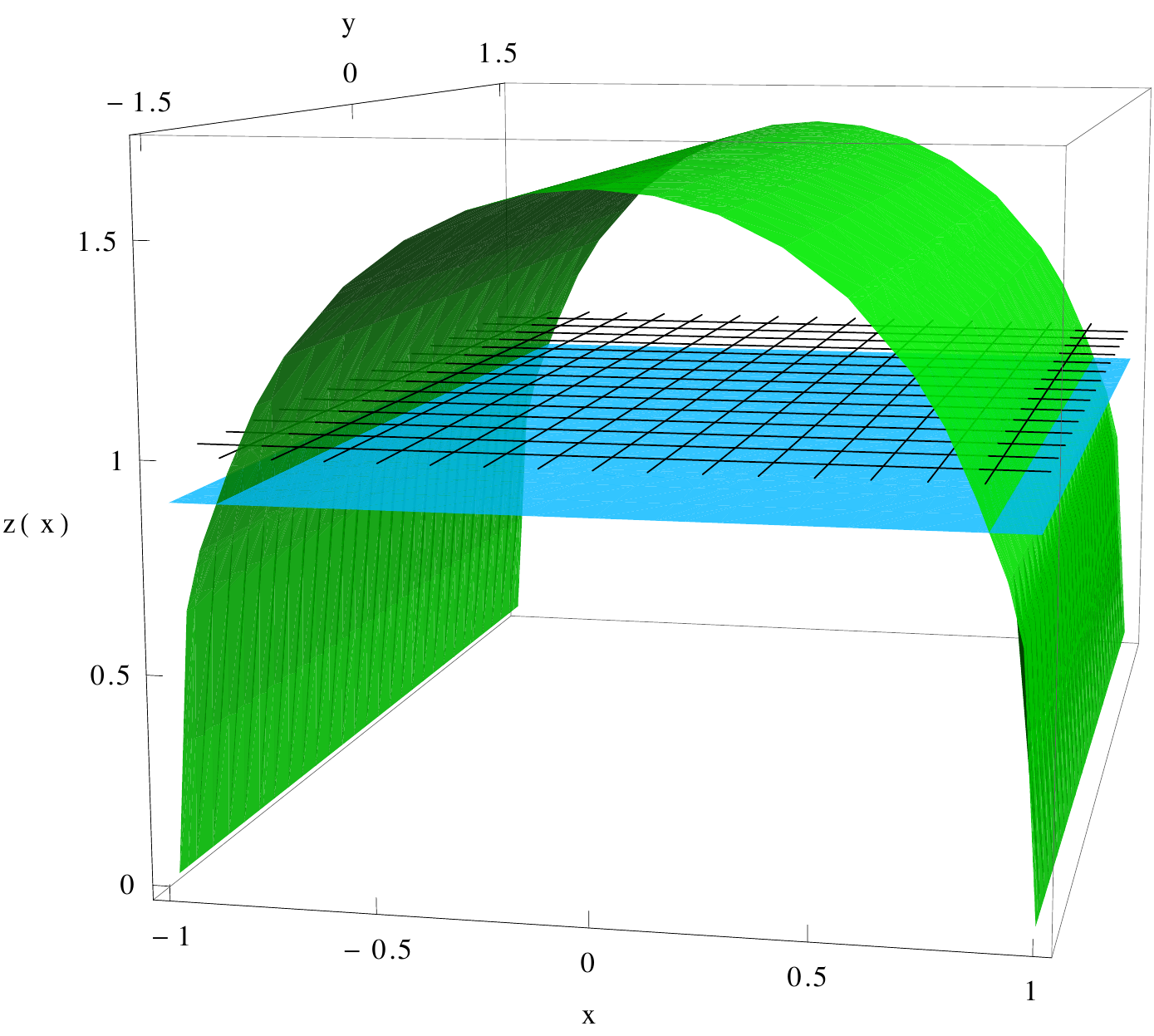}}\\
    \subfigure[$b=0$ at $t=1.54$]{\includegraphics[height=1.4in]{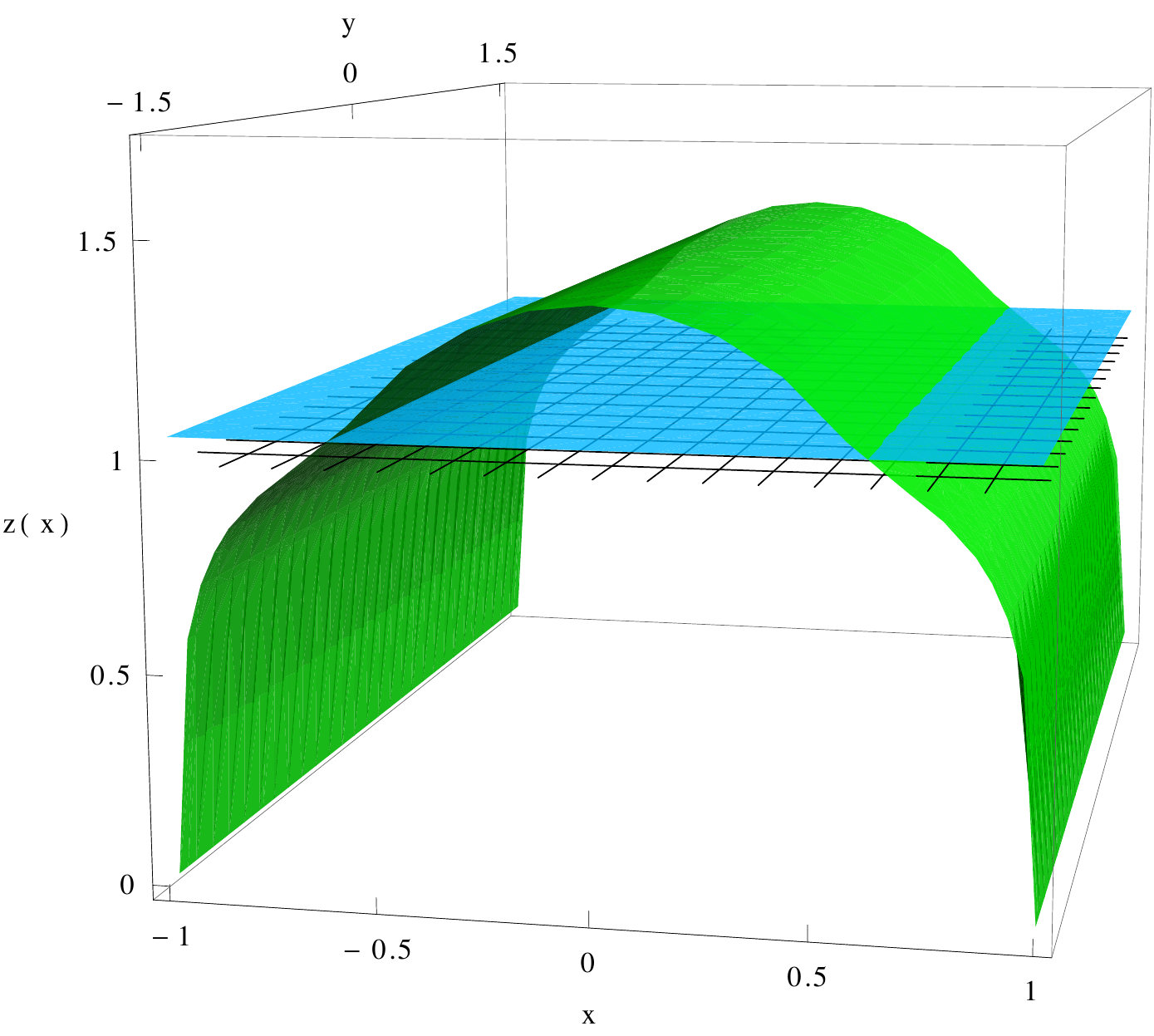}}\qquad
    \subfigure[$b=0.4$ at $t=1.54$]{\includegraphics[height=1.4in]{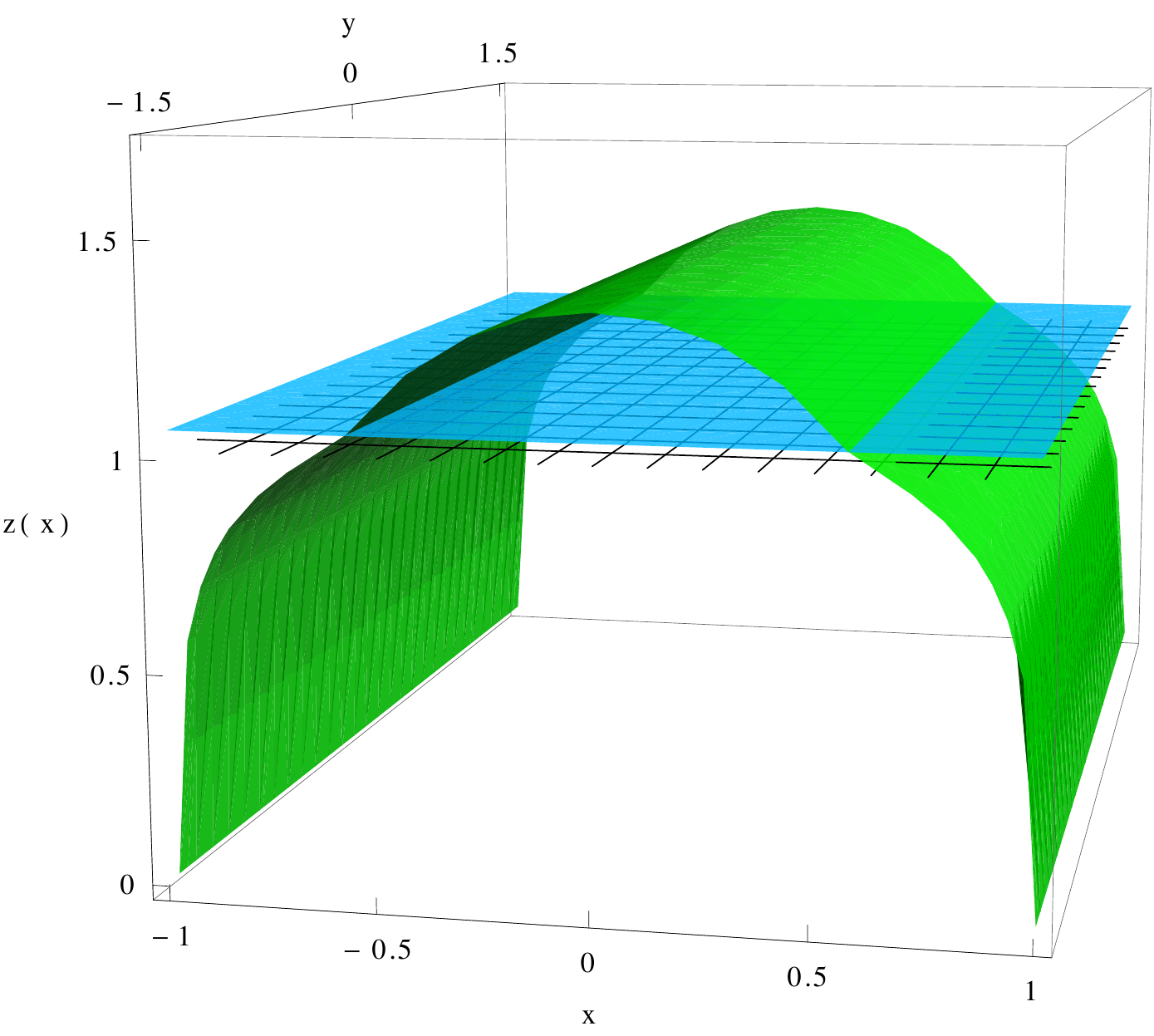}}\qquad
    \subfigure[$b=1$ at $t=1.54$]{\includegraphics[height=1.4in]{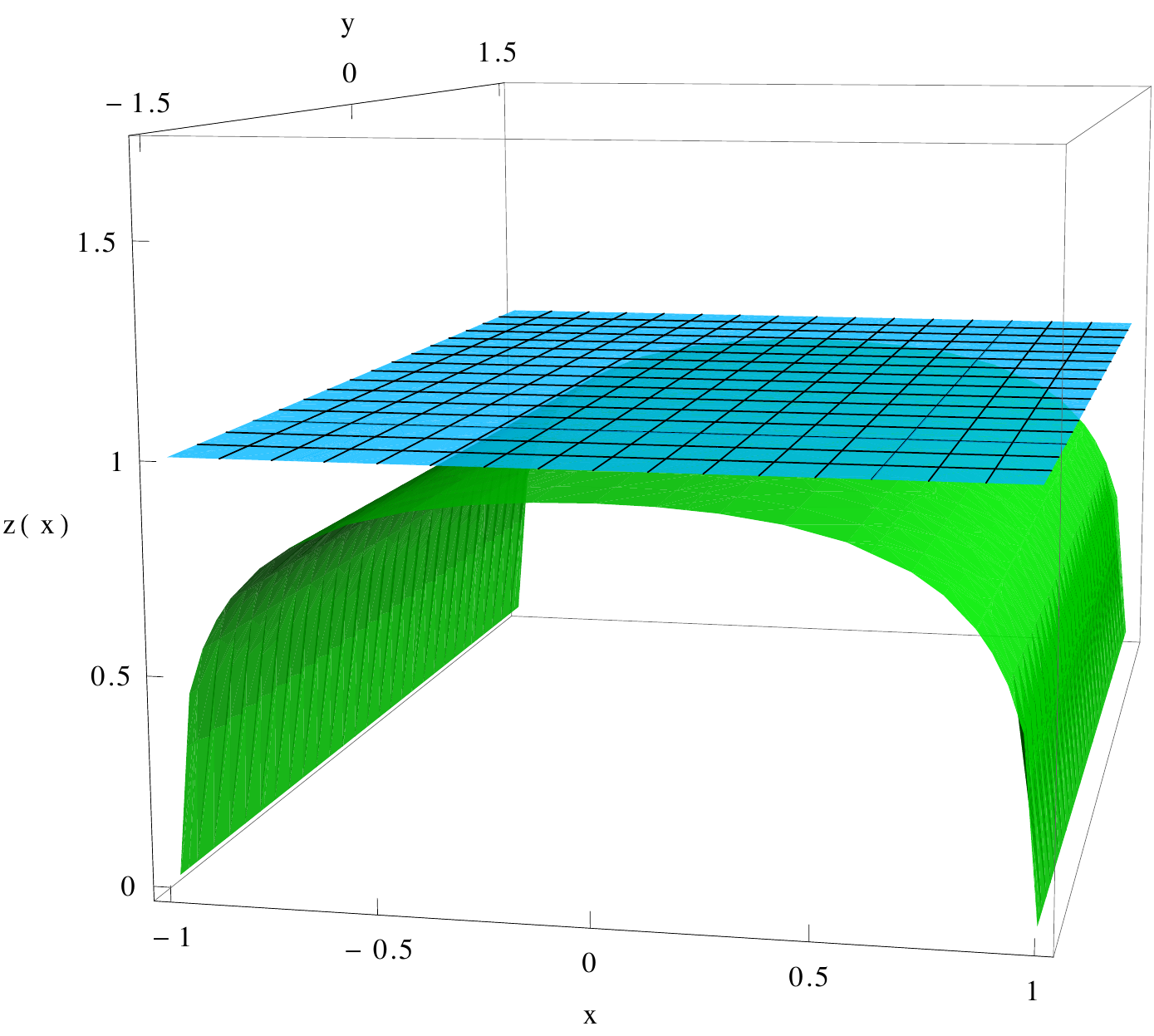}}\\
    \caption{A sequence of snapshots of the time evolution of the minimal area surface and the shell of charged dust described by the Vaidya-BI-AdS metric for different values of the inverse BI parameter $b$ and fixed charge $Q=\sqrt{2}$. In all cases the boundary Wilson loop has sides $\ell=2$ in the $x$ direction and $R=3$ in the $y$ direction. Each column, from top to bottom, indicates the time evolution for a given value of $b$. The cyan plane indicates the position of the shell in each case, while the gridded plane at $z=1$ denotes the position of the (still to be formed at late times) black brane horizon.}\label{fig:areaprofile}
\end{figure}

A sequence of snapshots of the time evolution of the minimal area surfaces as well as the shell of charged dust described by the Vaidya-BI-AdS metric is shown in Figure \ref{fig:areaprofile} for different values of the inverse BI parameter $b$ and a fixed charge $Q=\sqrt{2}$. Each column, from top to bottom, follows the time evolution for a given value of $b$. Again we can see that up to $t\sim1.0$ the value of $b$ has little effect on the dynamics, while at the final stages of the evolution $b$ plays a decisive role.
This is clear from the bottom row at $t=1.54$, where we see that the $b=1$ black brane has already formed while the other two are about to form. That illustration suggests that increasing the value of $b$ makes the black hole form earlier, which indeed will be confirmed below.

\begin{figure}[htbp]% Thermalization at b fixed
    \centering
    \subfigure[$b=0$]{\includegraphics[height=1.24in]{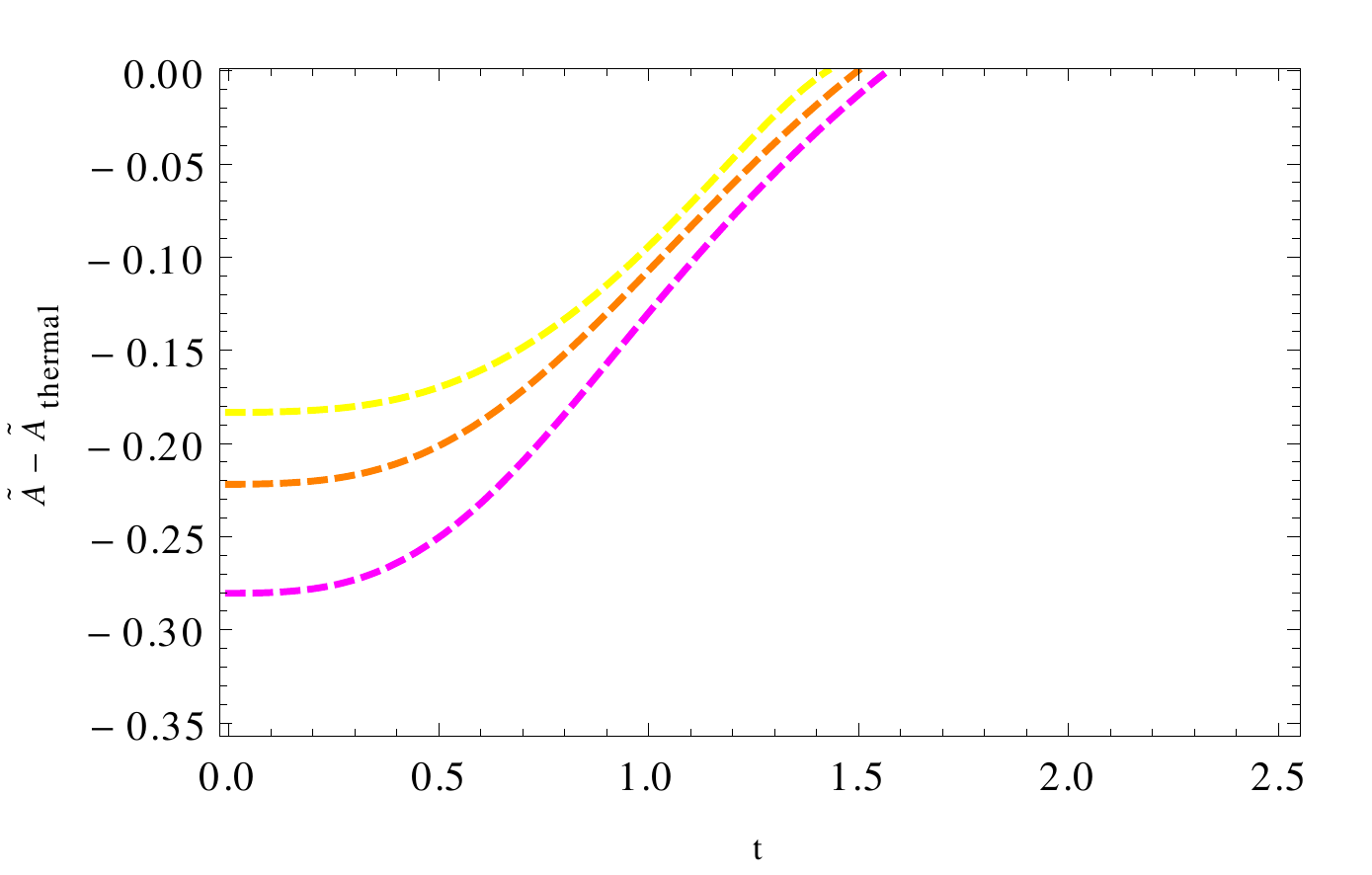}}
    \subfigure[$b=0.4$]{\includegraphics[height=1.24in]{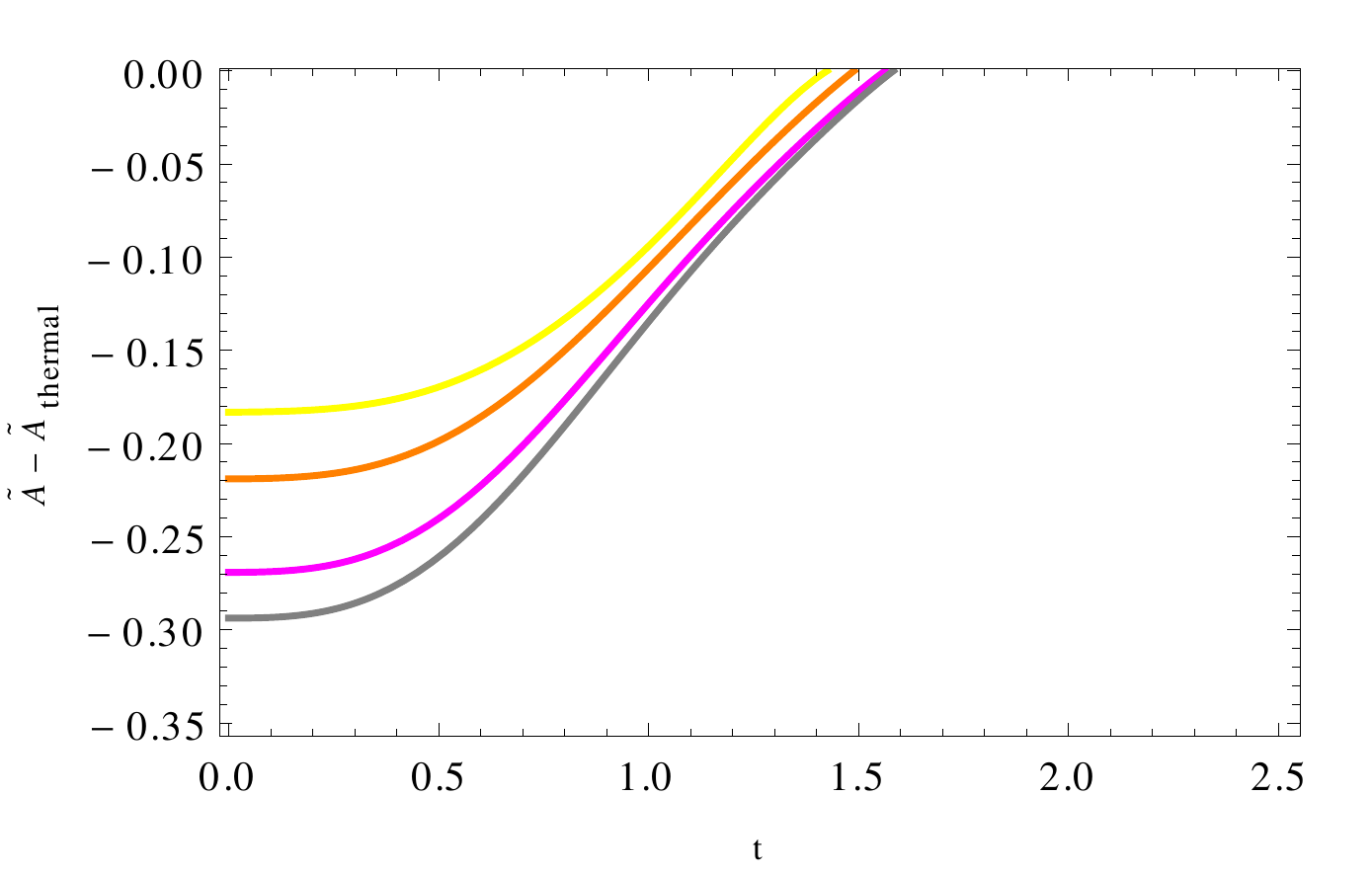}}
    \subfigure[$b=1$]{\includegraphics[height=1.24in]{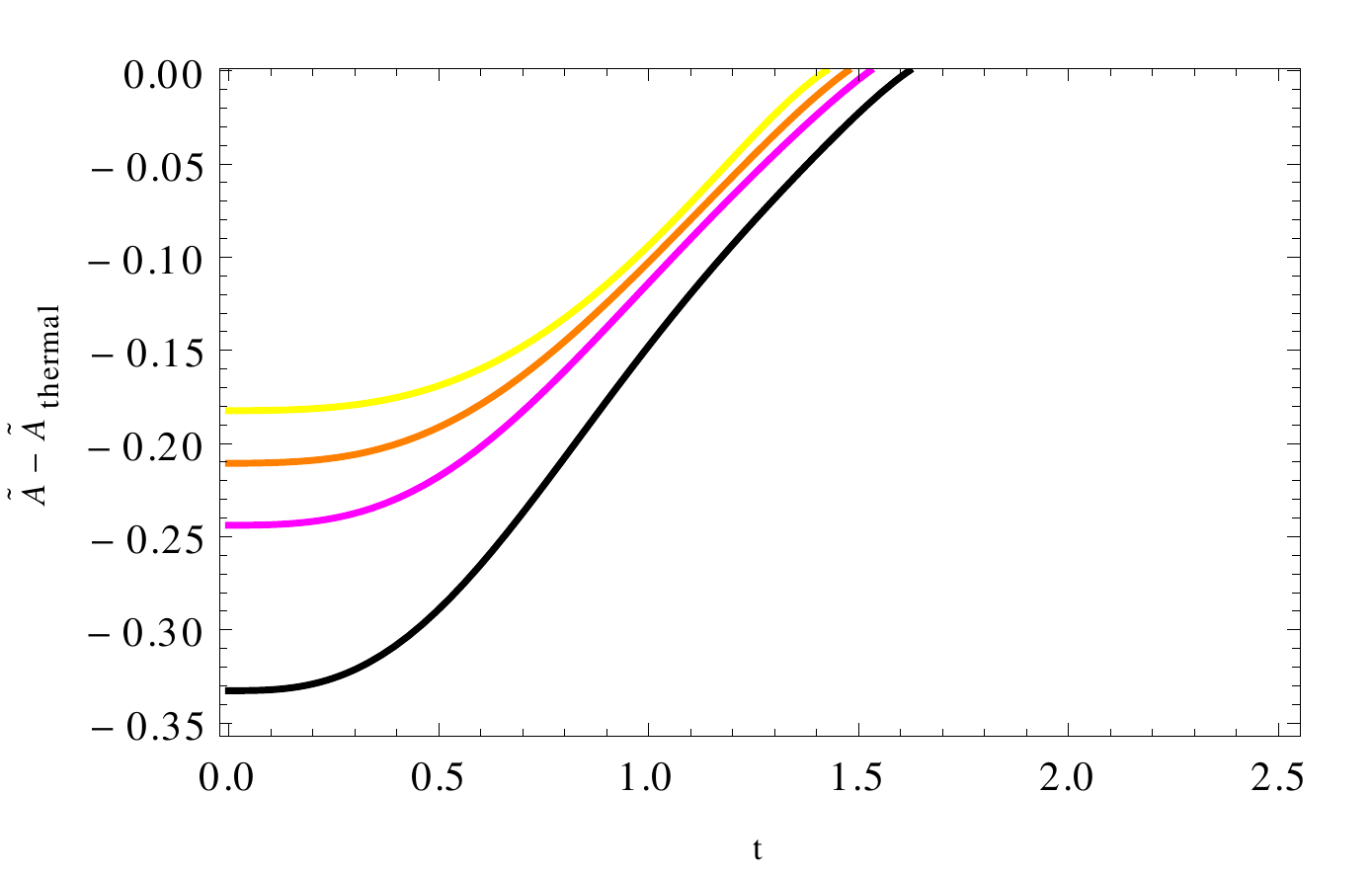}}\\
    \caption{Thermalization curves of the renormalized minimal area surfaces in the Vaidya-BI-AdS spacetime at fixed values of $b$ for different charges $Q$, from $Q=0.5$ to the extremal charge $Q_{ext}(b)=\sqrt{2+3b^2}$. In (a) we show $Q=0.5$ (yellow) in the top, $Q=1$ (orange) in the middle, and $Q=\sqrt{2}$ (magenta) in the bottom. In (b) and (c) the same values appear together with the extremal values $Q=\sqrt{2.48}$ (gray) and $Q=\sqrt{5}$ (black), respectively. The relevant side of the boundary Wilson loop is $\ell=2$.}\label{fig:areabfixed}
\end{figure}

\begin{figure}[htbp]% Thermalization at Q fixed
    \centering
    \subfigure[$Q=0.5$]{\includegraphics[height=1.24in]{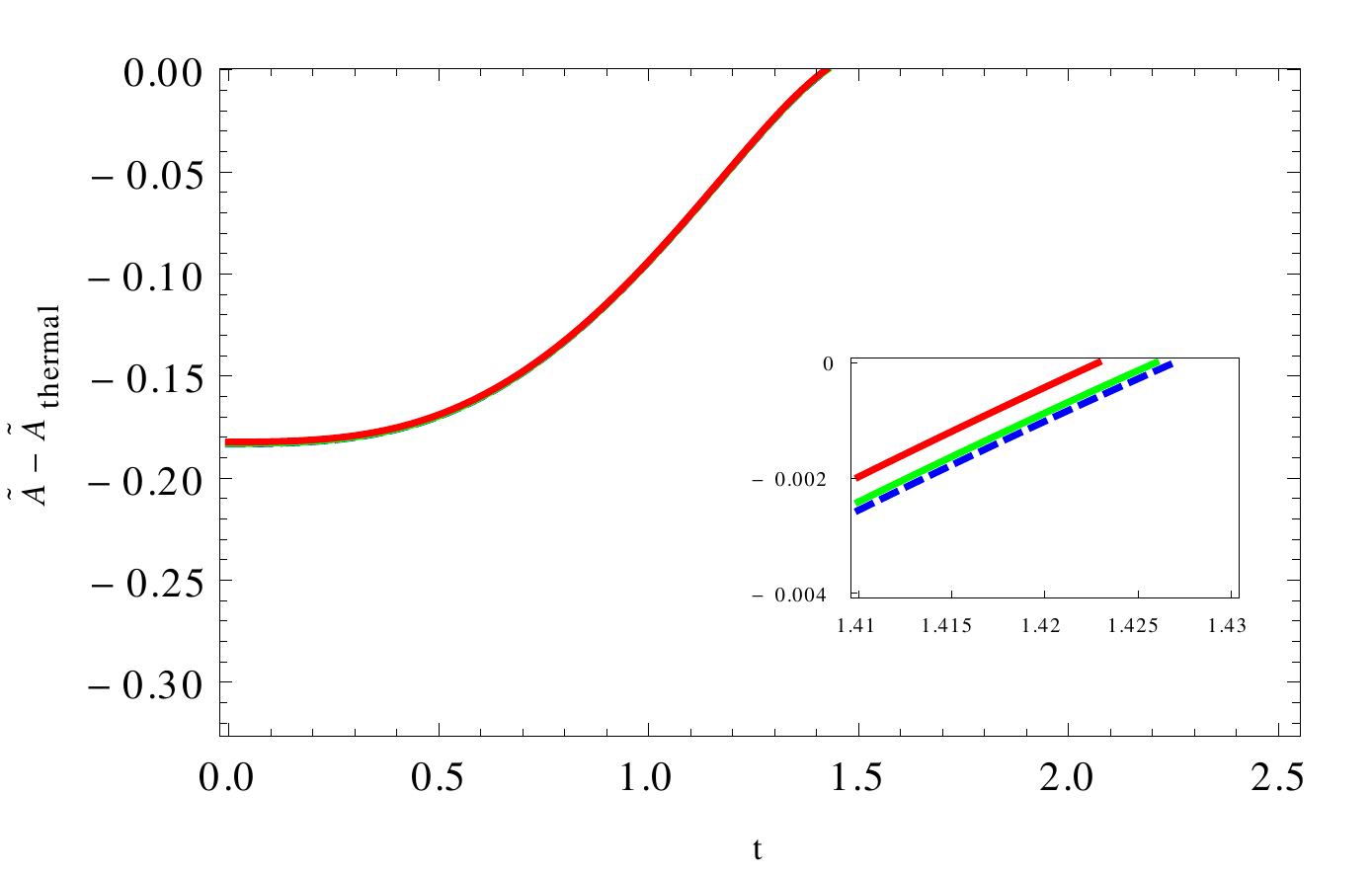}}
    \subfigure[$Q=1$]{\includegraphics[height=1.24in]{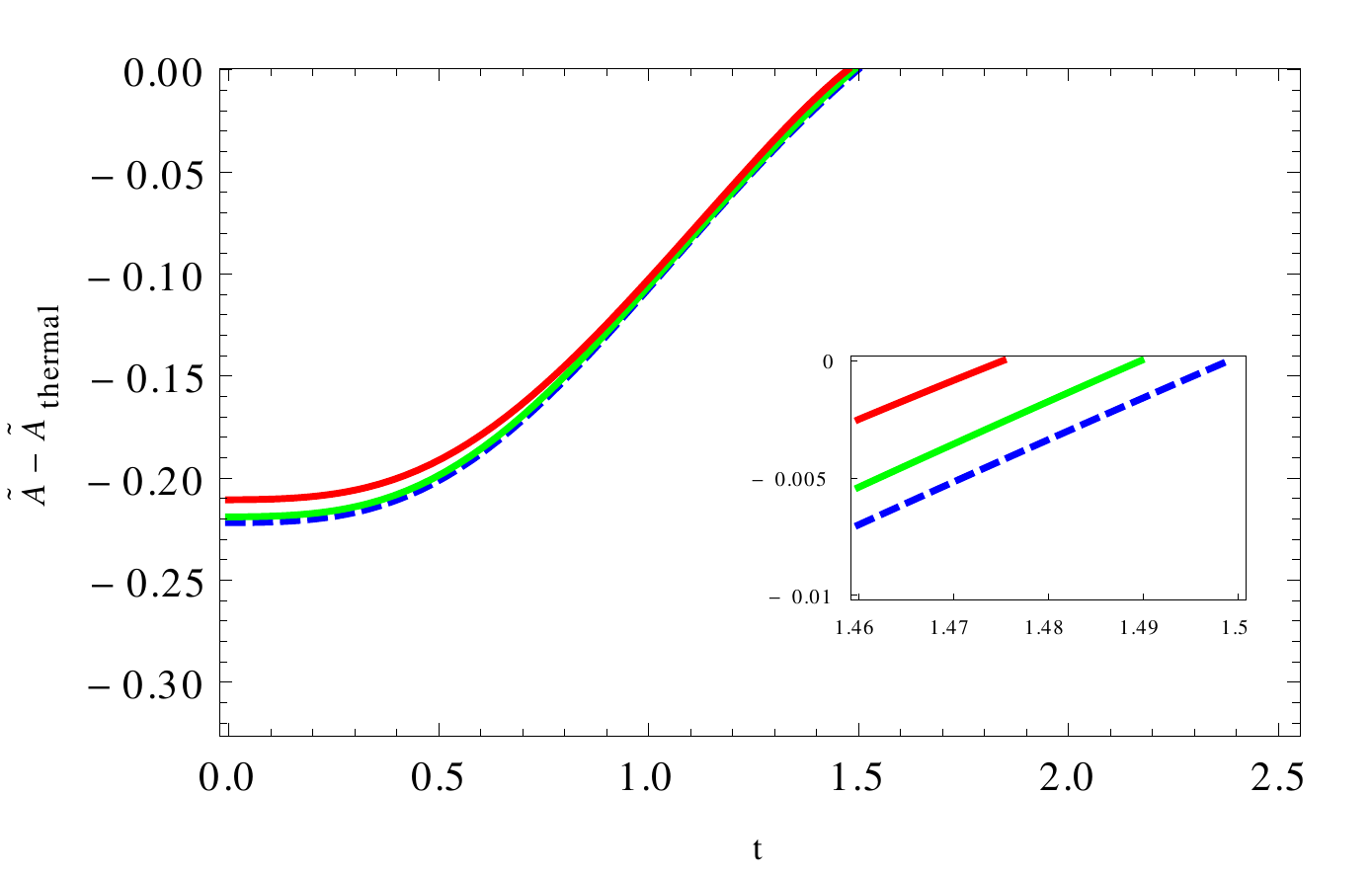}}
    \subfigure[$Q=\sqrt{2}$]{\includegraphics[height=1.24in]{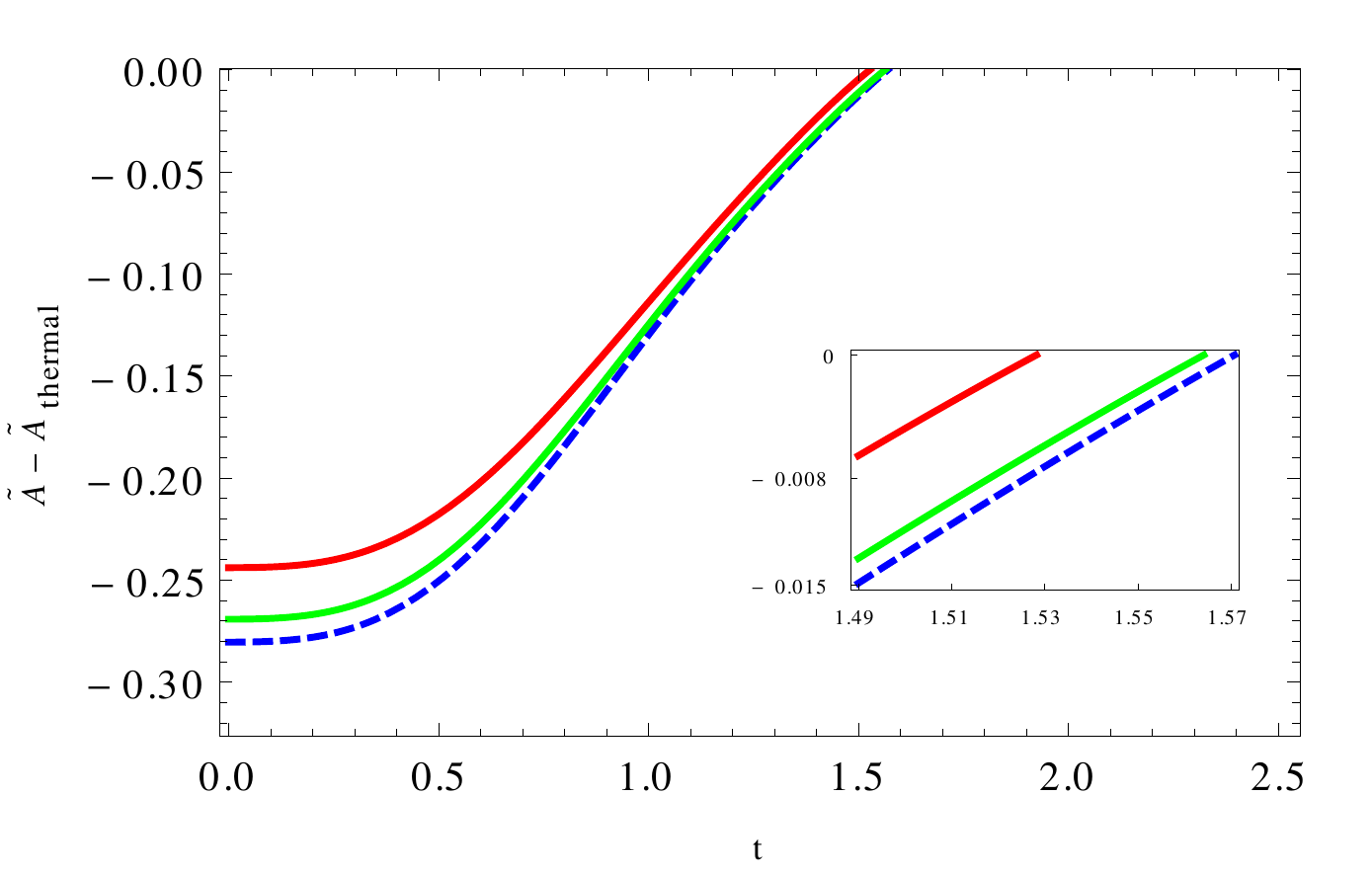}}\\
    \caption{Thermalization curves of the renormalized minimal area surfaces in the Vaidya-BI-AdS spacetime for different inverse BI parameters $b$ at fixed values of charge. In each case we have $b=0$ (dashed blue) in the bottom, $b=0.4$ (green) in the middle, and $b=1$ (red) in the top. The insets show details of the curves right before thermalization. The side of the boundary Wilson loop is $\ell=2$.}\label{fig:areaQfixed}
\end{figure}

The thermalization curves for the renormalized minimal area surfaces are shown in Figures \ref{fig:areabfixed} and \ref{fig:areaQfixed} for varying $Q$ at fixed values of $b$ and varying $b$ at fixed charges, respectively. All the curves are polynomial fits of the numerical points (see below) and the zero point of the vertical axis corresponds to the final state of the process, i.e., the static Einstein-BI black brane fully formed. We immediately notice that all the effects are less evident than those displayed before for the geodesic lengths (the reason why we show the insets in Figure \ref{fig:areaQfixed}) due to our choice of $\ell=2$ as the characteristic scale in the boundary, in contrast to the $\ell=4$ used for the geodesics. This just illustrates the argument made in the beginning that the holographic thermalization is a top-down process. Figure \ref{fig:areabfixed} shows that for a given $b$ the effect of the charge $Q$ is to delay the  thermalization process as it is increased, that is to say, as the chemical potential in the dual field theory grows, the time needed to reach the thermal state also raises. This reinforces the conclusions drawn from the analysis of the renormalized geodesic lengths in the previous subsection. The effect of the inverse BI parameter $b$ for fixed charges can be inferred from Figure \ref{fig:areaQfixed}, namely, the larger $b$ is, the shorter the thermalization time is. This implies that the boundary field theory is easier to thermalize in the nonlinear case, which is in perfect agreement with our results from the previous subsection. The numerical values obtained for the thermalization times are summarized in Table \ref{tab:areatimes}.

\begin{table}[htbp]
\centering
\begin{tabular}{c|c|c|c|c}
\cline{2-4}
                            & $b=0$   & $b=0.4$ & $b=1$   &  \\ \cline{1-4}
\multicolumn{1}{|c|}{$Q=0.5$} & $1.427$ & $1.426$ & $1.423$ &  \\ \cline{1-4}
\multicolumn{1}{|c|}{$Q=1$}   & $1.499$ & $1.490$ & $1.475$ &  \\ \cline{1-4}
\multicolumn{1}{|c|}{$Q=\sqrt{2}$}   & $1.571$ & $1.564$ & $1.528$ &  \\ \cline{1-4}
\multicolumn{1}{|c|}{$Q=\sqrt{2.48}$}   & -- & $1.584$ & -- &  \\ \cline{1-4}
\multicolumn{1}{|c|}{$Q=\sqrt{5}$}   & -- & -- & $1.622$ &  \\ \cline{1-4}
\end{tabular}
\caption{Summary of the numerical values obtained for the thermalization times of the renormalized minimal area surfaces shown in Figures \ref{fig:areabfixed} and \ref{fig:areaQfixed}.}\label{tab:areatimes}
\end{table}

\begin{figure}[htbp]% Thermalization velocities
    \centering
    \subfigure[$Q=\sqrt{2}$]{\includegraphics[height=1.35in]{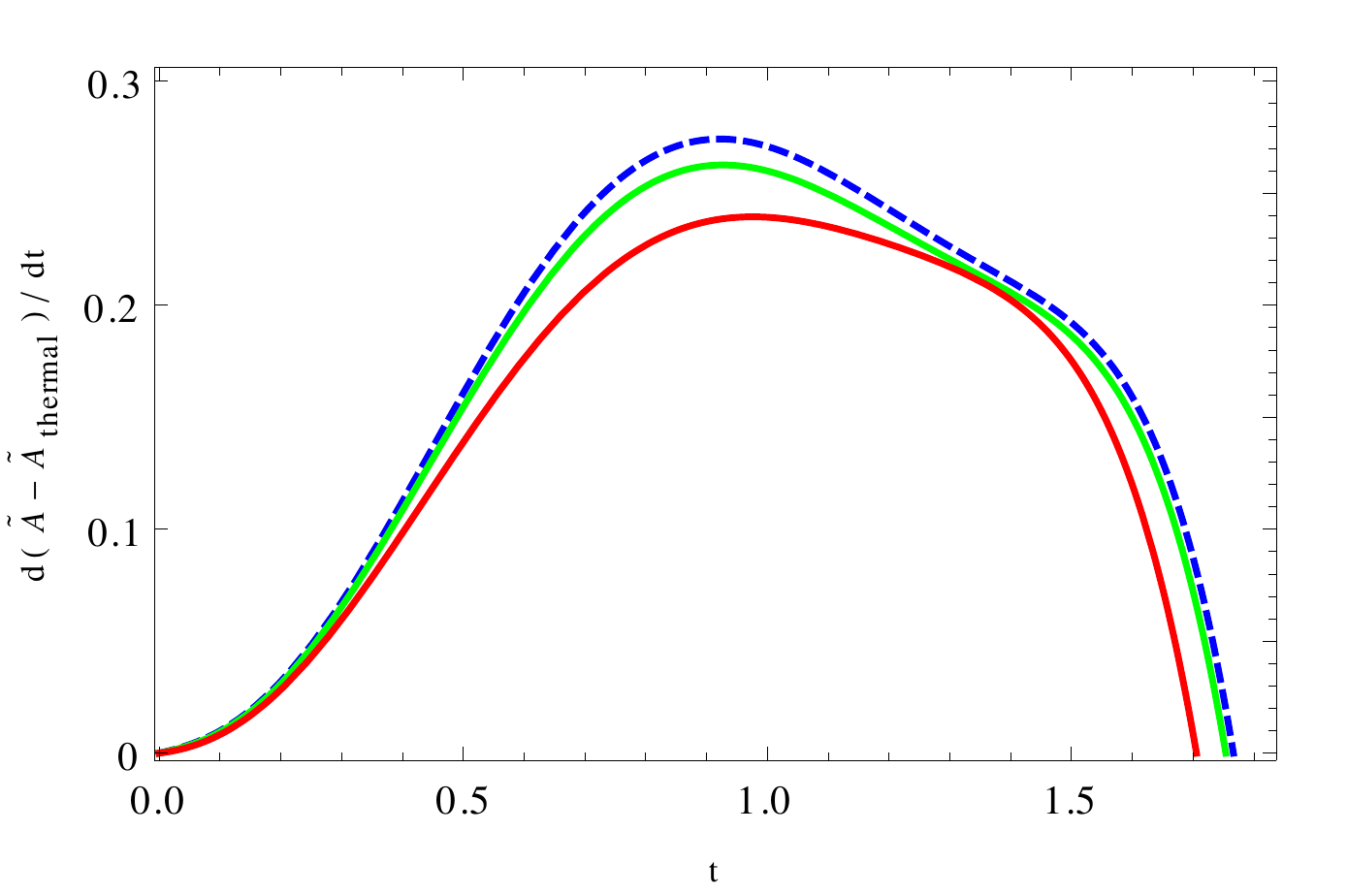}}\label{fig:veloareaQext}\qquad
    \subfigure[$b=1$]{\includegraphics[height=1.35in]{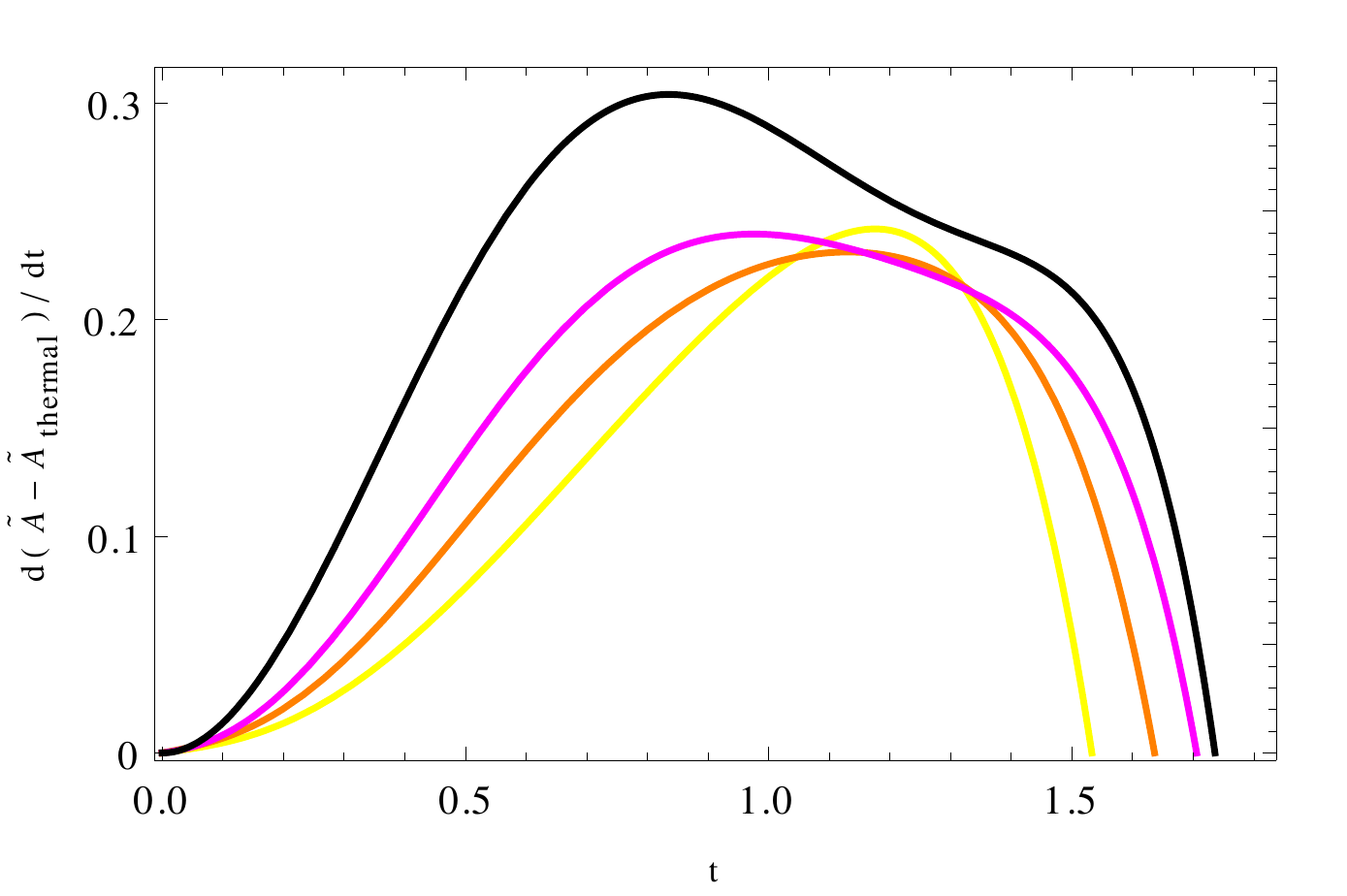}}\label{fig:veloareab1}
    \caption{Thermalization velocities of the renormalized minimal area surfaces at both fixed charge and fixed $b$. The dashed blue, green, and red curves in (a) correspond respectively to $b=0,0.4,1$. The curves in (b) correspond to $Q=0.5$ (yellow), $Q=1$ (orange), $Q=\sqrt{2}$ (magenta), and $Q=\sqrt{5}$ (black). The side of the boundary Wilson loop is $\ell=2$.}\label{fig:areavelo}
\end{figure}

Just as we did before, we also use the fit functions to study the thermalization velocities for the minimal area surfaces, $\frac{d}{dt}(\tilde{\mathcal{A}}-\tilde{\mathcal{A}}_{\textrm{thermal}})$, which are plotted in Figure \ref{fig:areavelo} for the cases $Q=\sqrt{2}$ fixed and $b=1$ fixed, respectively. The plots confirm the results obtained from the geodesics in what concerns the different stages of the dynamical process. Namely, there is always a phase transition instant at the middle stage of the evolution where the process changes from an accelerating phase to a decelerating phase. Such a phase transition point is reached later as we increase $b$ for a fixed charge (see Figure \ref{fig:veloareaQext}), or sooner as the charge $Q$ is increased for a given $b$ (see Figure \ref{fig:veloareab1}). In other words, the thermalization of expectation values of Wilson loops in the dual field theory consists in a slow (quick) accelerating phase followed by a quick (slow) decelerating phase towards the final state as the nonlinearity parameter $b$ (the charge $Q$, or chemical potential $\mu/T$) is increased.

\begin{figure}[thbp]% Comparation data/fits
    \centering
    \subfigure[$Q=1,b=0$]{\includegraphics[height=1.35in]{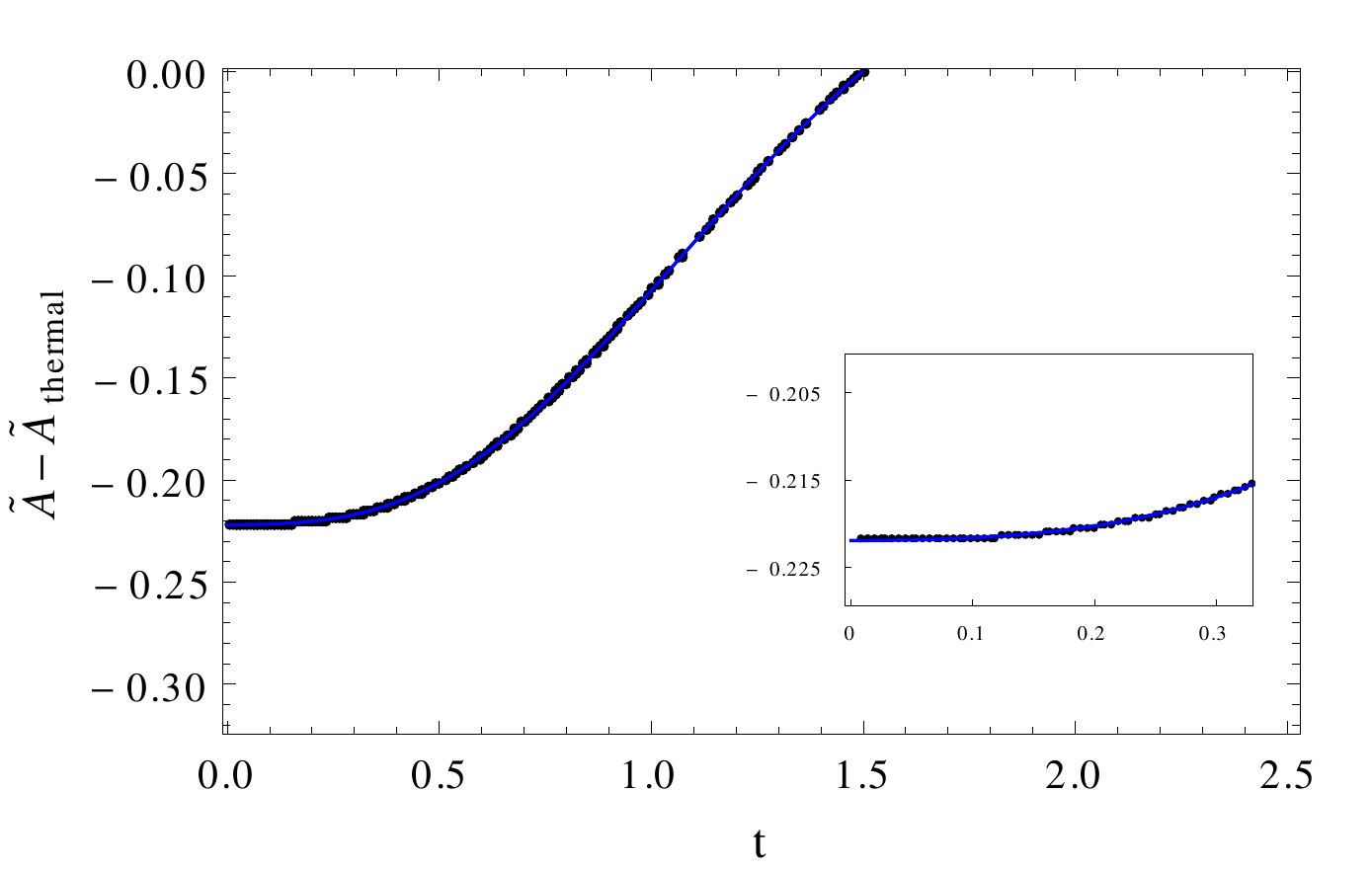}}\qquad
    \subfigure[$Q=\sqrt{2},b=1$]{\includegraphics[height=1.35in]{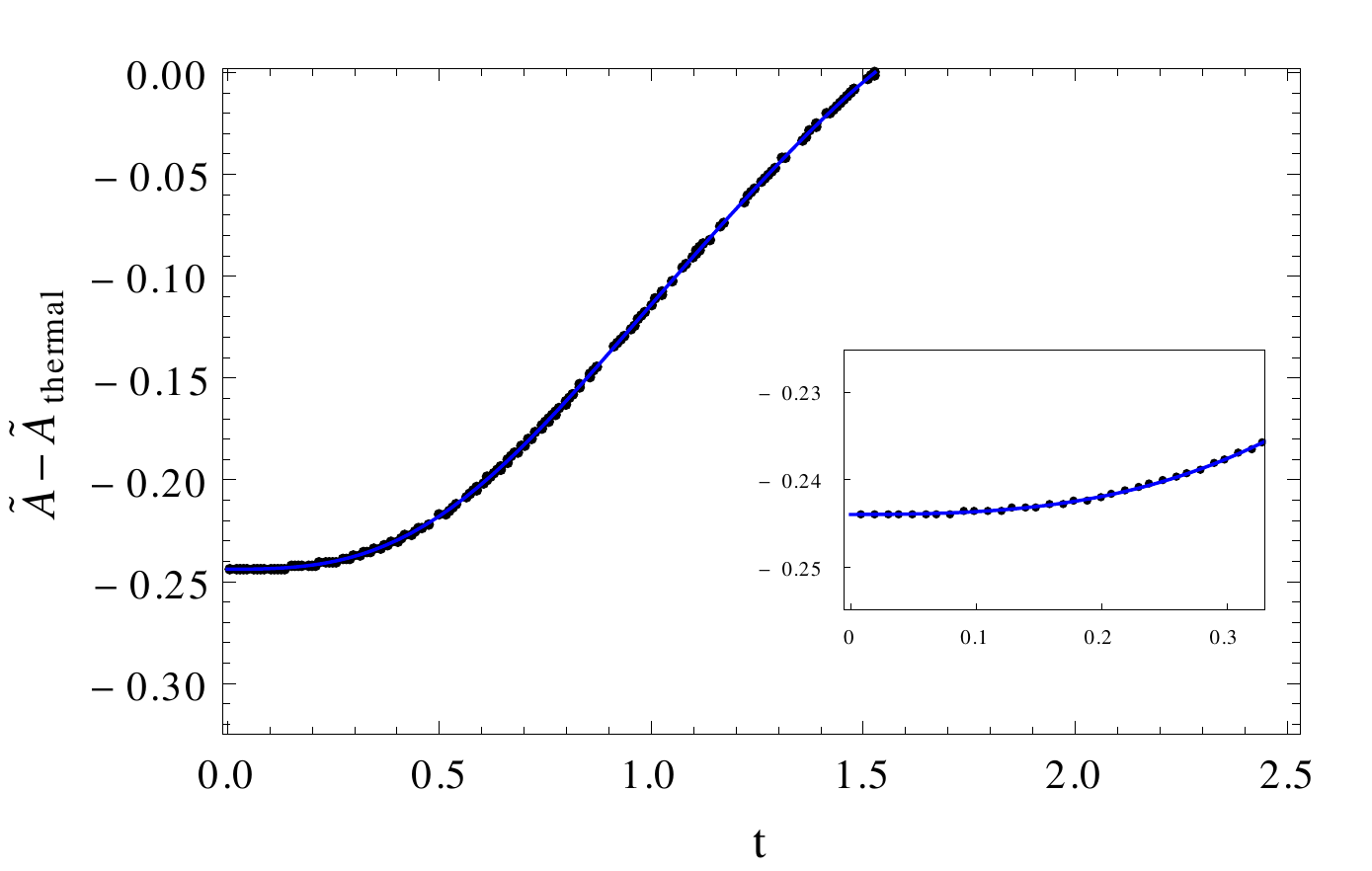}}
    \caption{A comparison between the numerical results and the polynomial fits for the minimal area surfaces. The inset emphasizes the constant behavior of the curve at initial times.}\label{fig:areafits}
\end{figure}

Once more we point out that our velocity plots do not provide any evidence of a negative thermalization velocity at initial times, in contrast to \cite{chinesesGBQ}. This can also be seen from Figure \ref{fig:areafits}, where we show a comparison of the numerical results with the fitting functions. The zoomed region shows that at the initial times, up to $t\sim0.1$, all the numerical points lie over a horizontal line and thus the curve must have vanishing slope in that region. We should mention that the fitting functions used here were degree $7$ polynomials $f(t)=\sum_{n=0}^{7}\alpha_nt^n$ with the linear term ($\alpha_1$) set to zero to ensure the strictly constant behavior $f(t)\sim \alpha_0$ up to $t\sim0.1$ as required by the numerical data.

%%%%%%%%%%%%%%%%%%%%%%%%%%%%%%%%%%%%%%%%%
\section{Conclusions}\label{conclusions}
%%%%%%%%%%%%%%%%%%%%%%%%%%%%%%%%%%%%%%%%%

The effect of the chemical potential and the inverse BI parameter on the thermalization time of the dual boundary field theory is studied using the Vaidya-like toy model of a collapsing shell of charged dust. The bulk spacetime is dynamical, constructed to interpolate between a pure AdS space at initial times and a charged Einstein-Born-Infeld AdS black brane at late times. We use as thermalization probes the equal-time two-point functions and expectation values of Wilson loops, which have well defined dual gravity descriptions in terms of renormalized geodesic lengths and minimal area surfaces in the bulk. Another class of nonlocal observables, the entanglement entropy of boundary regions, which also have well known holographic descriptions in terms of minimal volumes of codimension-2 surfaces in the bulk, can also be used as a third thermalization probe. However, as the results are similar, we focus only on the first two since they are enough to capture all the relevant effects. 

We conclude that as the charge (or, equivalently, the chemical potential) grows, the thermalization time is also increased. The inverse BI parameter, on the other hand, has the opposite effect, i.e., the larger is the value of $b$, the shorter the thermalization time is. At the initial stages of the thermalization we find that $Q$ and $b$ have little effect on the time evolution, becoming important only from the middle stage on. We arrive at the same results independently for both the renormalized geodesic lengths and minimal area surfaces. In each case they can be seen just by looking to a sequence of snapshots of the motion profiles of the geodesics and minimal surfaces as the time goes by or, alternatively, from the thermalization curves obtained from the full numerical analysis. The effect of the charge happens to be the same found in \cite{galante} using Einstein-Maxwell theory and \cite{chinesesGBQ}, where Gauss-Bonnet curvature corrections were included. The outcome of introducing a nonvanishing $b$, on the other hand, is a new result. In particular, the effect is the same as that of the Gauss-Bonnet parameter reported in \cite{chinesesGBQ}. Since BI electrodynamics consists essentially of higher order derivatives of the gauge field, we suggest that this may be a general feature of introducing extra derivatives in the bulk, i.e., that boundary gauge theories whose gravity dual carry more than two derivatives tend to thermalize sooner than two-derivative theories after a sudden injection of energy. It would be interesting to explore this idea with other theories possessing this characteristic in order to have definite answers.

Moreover, by fitting the numerical data with smooth functions we were also able to study the thermalization velocities associated to each curve. Although this is not rigorous, it reveals some interesting features of the dynamical process of thermalization and how they are affected by $Q$ and $b$. We notice the existence of a phase transition point at the middle stage of the thermalization, which divides the process into an accelerating and a decelerating phase. The phase transition point is shifted depending on the values of $b$ and $Q$. Namely, as $Q$ increases, the phase transition point arrives earlier. This indicates that for large values of chemical potential the thermalization process consists of a quick accelerating phase followed by a slowly decelerating phase to the final state. Increasing the value of $b$, oppositely, causes a delay in the phase transition point. This is to be contrasted with the fact that nonlinear theories thermalize first, indicating that the dynamical process for non-vanishing $b$ consists of a slowly accelerating phase followed by a quick deceleration towards the equilibrium state. We also show from the velocity plots that the thermalization process is monotonic and the velocity is always positive at the initial stages, in contrast to the claim by the authors of \cite{chinesesGBQ} that there should be a negative thermalization velocity at the very beginning of the evolution.

\vspace{1cm}

%%%%%%%%%%%%%%%%%%%%%%%%%%%%%%%%%%%%%%%%%%
\centerline{\large{\bf Acknowledgments}}
%%%%%%%%%%%%%%%%%%%%%%%%%%%%%%%%%%%%%%%%%%

~

We would like to thank Gast\'on Giribet for valuable discussions at the early stages of this work. G. Camilo and E. Abdalla thank CNPq for financial support. E. Abdalla also thanks the support from FAPESP.

\end{document}